\documentclass[a4paper,fleqn,usenatbib]{mnras}

\usepackage{newtxtext,newtxmath}
\usepackage[T1]{fontenc}

\usepackage[english]{babel}
\usepackage{graphicx}
\usepackage{fleqn}
\usepackage{amssymb}
\usepackage{amsmath}
\usepackage{booktabs}
\usepackage{hyperref}
\usepackage{comment}
\usepackage{enumitem}
\setlist[enumerate]{
  labelsep=8pt,
  labelindent=0.\parindent,
 itemindent=0pt,
  leftmargin=*,
}

\usepackage{color, colortbl}

\newcommand*{\mycdot}{\kern-.2em\cdot\kern-.2em}
\renewcommand{\S}{Section}
\newcommand{\F}{Fig.}
\newcommand{\codename}{SecularMultiple}

\newcommand{\ve}[1]{\boldsymbol{#1}}
\newcommand{\unit}[1]{\hat{\boldsymbol{#1}}}
\newcommand{\ma}[1]{\mathrm{\bf{#1}}}

\newcommand{\complexi}{\mathrm{i}}

\newcommand{\msun}{\mathrm{M}_\odot}

\newcommand{\au}{\,\textsc{au}}
\newcommand{\renc}{R_\mathrm{enc}}

\newcommand{\hpz}{Paper I}

\pubyear{2018}

\usepackage{listings}

\usepackage{color}

\definecolor{dkgreen}{rgb}{0,0.6,0}
\definecolor{gray}{rgb}{0.5,0.5,0.5}
\definecolor{mauve}{rgb}{0.58,0,0.82}

\allowdisplaybreaks

\begin{document}

\title[Perturbed hierarchical multiple systems]{Secular dynamics of hierarchical multiple systems composed of nested binaries, with an arbitrary number of bodies and arbitrary hierarchical structure. II. External perturbations: flybys and supernovae}

\author[Hamers]{Adrian S. Hamers$^{1}$\thanks{E-mail: hamers@ias.edu} \\
$^{1}$Institute for Advanced Study, School of Natural Sciences, Einstein Drive, Princeton, NJ 08540, USA}
\date{Accepted 2018 February 14. Received 2018 January 24; in original form 2017 December 11.}

\label{firstpage}
\pagerange{\pageref{firstpage}--\pageref{lastpage}}
\maketitle

\begin{abstract} 
We extend the formalism of a previous paper to include the effects of flybys and instantaneous perturbations such as supernovae on the long-term secular evolution of hierarchical multiple systems with an arbitrary number of bodies and hierarchy, provided that the system is composed of nested binary orbits. To model secular encounters, we expand the Hamiltonian in terms of the ratio of the separation of the perturber with respect to the barycentre of the multiple system, to the separation of the widest orbit. Subsequently, we integrate over the perturber orbit numerically or analytically. We verify our method for secular encounters, and illustrate it with an example. Furthermore, we describe a method to compute instantaneous orbital changes to multiple systems, such as asymmetric supernovae and impulsive encounters. The secular code, with implementation of the extensions described in this paper, is publicly available within \textsc{AMUSE}, and we provide a number of simple example scripts to illustrate its usage for secular and impulsive encounters, and asymmetric supernovae. The extensions presented in this paper are a next step toward efficiently modeling the evolution of complex multiple systems embedded in star clusters. 
\end{abstract}


\section{Introduction}
\label{sect:introduction}
Astrophysical systems tend to be arranged in hierarchical configurations. Examples include multistar systems, and multiplanet systems around single and multiple stars. With the increasing number of high-multiplicity stellar systems known (e.g., \citealt{1997A&AS..124...75T,2014AJ....147...86T,2014AJ....147...87T}) and with the increasing number of exoplanets found, understanding the dynamical evolution of these systems is becoming increasingly important. At the same time, these systems do not live in isolation. In fact, most stars, if not all, are not born alone \citep{2003ARA&A..41...57L}, and even stars in the field that have long dissociated from their birth cluster still experience occasional flybys with other stars. 

Flybys can have important implications for the evolution of stellar and planetary systems. In the field, stellar encounters can drive wide binaries to high eccentricities, triggering strong tidal interactions and/or stellar collisions \citep{2014ApJ...782...60K}, destabilizing planetary systems \citep{2013Natur.493..381K}, or producing low-mass X-ray binaries \citep{2016MNRAS.458.4188M,2017MNRAS.469.3088K}. Similarly, flybys can excite comets in the Oort cloud, bringing them into the inner Solar system (\citealt{1950BAN....11...91O}, and, e.g, \citealt{1987Icar...70..269H,1987AJ.....94.1330D,2002A&A...396..283D,2002AN....323...37S,2011Icar..214..334F,2015AJ....150...26H}), or, in the case of exo-Oort clouds around evolved stars, trigger pollution of white dwarfs \citep{2014MNRAS.445.4175V}.

In the context of the evolution of multiplanet systems, stellar encounters can perturb the orbits of wide planets. Subsequently, these perturbations can be propagated inwards through planet-planet scattering or secular interactions (e.g., \citealt{2004AJ....128..869Z,2011MNRAS.411..859M,2012ApJ...754...57B,2013ApJ...772..142L,2013MNRAS.433..867H,2015MNRAS.453.2759Z,2015MNRAS.449.3543W,2017MNRAS.470.4337C}). Through this process, planets on close orbits can be strongly perturbed by distant flybys that, in the absence of planetary companions, would not affect tightly-bound planets.

Although $N$-body integrations offer the most straightforward and accurate method to address the dynamics of hierarchical multiple systems with the inclusion of flybys, they are computationally expensive, and do not allow for much insight into the fundamental physics. In a previous paper, \citeauthor{2016MNRAS.459.2827H} (\citeyear{2016MNRAS.459.2827H}; hereafter \hpz), we presented a formalism and an algorithm to compute the long-term secular evolution of isolated hierarchical multiple systems with any number of bodies and hierarchy, provided that they are composed of nested binary orbits. In a sense, this work constituted a generalization of the seminal works of \citet{1962P&SS....9..719L,1962AJ.....67..591K} for hierarchical triple systems. Lidov-Kozai (LK) oscillations in triple systems have important implications for a large range of systems (see \citealt{2016ARA&A..54..441N} for a review), including, e.g., short-period binaries \citep{1979A&A....77..145M,1998MNRAS.300..292K,2001ApJ...562.1012E,2006Ap&SS.304...75E,2007ApJ...669.1298F,2014ApJ...793..137N}, and hot Jupiters (e.g., \citealt{2003ApJ...589..605W,2007ApJ...669.1298F,2011Natur.473..187N,2012ApJ...754L..36N,2015ApJ...799...27P,2016MNRAS.456.3671A,2016ApJ...829..132P,2017ApJ...835L..24H}).

Here, we expand on \hpz\, and include the effects of stellar flybys. Specifically, we introduce a method to model secular encounters, which allows for an efficient computation of these effects coupled with the `internal' secular evolution of the system, and derive the equations of motion to arbitrary expansion order for pairwise interactions. In a sense, our new contribution is a generalization of the work of \citet{1996MNRAS.282.1064H}, who considered secular encounters with binaries up to and including third order (`octupole order'). We remark that secular encounters are a sub type of more general few-body interactions, including binary-single scattering (e.g., \citealt{1983ApJ...268..319H,1983ApJ...268..342H,1991MNRAS.250..555H}), and scattering involving triple systems (e.g., \citealt{2012MNRAS.425.2369L,2015MNRAS.450.1724L,2016MNRAS.456.4219A})

In addition to stellar flybys, we include a different type of external perturbations. Specifically, we implement a module within our algorithm that can be used to compute the effects of instantaneous perturbations associated with changes of the masses, positions and velocities of the bodies in the system. Among the applications of this module are the calculation of the effects of asymmetric supernovae, and impulsive encounters. 

Supernovae (SNe) play an important role in the evolution of hierarchical multiple systems in which one or more of the stars are sufficiently massive to produce neutron stars and/or black holes. In addition to affecting the orbits through mass loss in the SN \citep{1961BAN....15..265B,1961BAN....15..291B}, there is compelling evidence that the newly-formed compact object also receives a kick velocity in a random direction \citep{1970SvA....13..562S,1970ApJ...160..979G,1997ApJ...483..399V}. Formulae for the orbital changes due to SNe in hierarchical triple systems were given by \citet{2012MNRAS.424.2914P}, although the expressions for the eccentricity changes were incomplete, and changes in the orbital orientations (in particular, the mutual inclination) were not considered. Here, we generalize the results of \citet{2012MNRAS.424.2914P} to compute asymmetric SNe changes in hierarchical triples to hierarchical multiple systems, and implement a routine in the secular code of \hpz, allowing to easily combine the short-term effects of SNe with long-term secular evolution. In addition, our implementation allows for any change in the masses, positions and velocities of any of the bodies, making it very general. For example, it can also be used to easily include the effects of impulsive encounters. 

Both implementations of stellar flybys and instantaneous perturbations presented in this paper are included within the code \textsc{\codename}, which is freely available and part of the \textsc{AMUSE} framework\footnote{At the time of submission of this manuscript, the updates presented here are not yet part of \textsc{\codename} in the official release of \textsc{AMUSE}. However, they are included in the \texttt{GitHub} version of \textsc{AMUSE} which is available at \href{https://github.com/amusecode/amuse}{https://github.com/amusecode/amuse}.} \citep{2013CoPhC.183..456P,2013A&A...557A..84P}. 

This paper is organized as follows. In \S\,\ref{sect:sec}, we describe our formalism to model secular encounters with hierarchical multiple systems. We verify the method, illustrate how to use the code in practice, and give a brief example of how encounters can affect the evolution of BH triples in globular clusters. In \S\,\ref{sect:inst}, we discuss our treatment of instantaneous changes to the system, and demonstrate its use for SNe and impulsive encounters. We briefly apply the treatment of impulsive encounters to the evolution of wide 2+2 quadruples in the Solar neighborhood. We discuss our results in \S\,\ref{sect:discussion} and conclude in \S\,\ref{sect:conclusions}. A detailed discussion on the secular encounter methodology is given in Appendix\,\ref{app:per}.

\section{Secular encounters}
\label{sect:sec}
\subsection{Formalism and main approximations}
\label{sect:sec:form}

\begin{table}
\begin{tabular}{lp{5.0cm}}
\toprule
Symbol & Description \\
\midrule
$G$						& Gravitational constant. \\
$H_\mathrm{int}$			& Hamiltonian associated with the internal system. \\
$H_\mathrm{per}$			& Hamiltonian associated with the perturber. \\
$n$						& Expansion order in the Hamiltonian. \\
$m_i$					& Mass of body $i$ in the internal system. \\
$M_p$					& Mass of binary $p$, i.e., sum of the masses of all bodies contained within binary $p$. \\
$M_\mathrm{per}$			& Mass of the perturber. \\
$\mathcal{M}_p^{(n)}$		& Dimensionless mass ratio associated with binary $p$ and expansion order $n$, defined in equation~(\ref{eq:mathcal_M}). \\
$M_{p.\mathrm{C}i}$ 		& The mass of all bodies contained within child $i$ of binary $p$. \\
$a_p$					& Semimajor axis of binary $p$. \\
$\tilde{n}_p$				& Mean motion of binary $p$. \\
$\tilde{n}_\mathrm{per}$		& Mean hyperbolic motion. \\
$\dot{f}_\mathrm{per}$		& Time derivative of the true anomaly of the perturber evaluated at periapsis. \\
$\mathcal{R}_w$			& Angular speed ratio (see equation~\ref{eq:R_def}) defined with respect to the widest orbit in the internal system (binary $w$). \\
$\ve{e}_p$				& Eccentricity vector of binary $p$. \\
$\ve{\jmath}_p$				& Normalized angular-momentum vector of binary $p$; its norm is $\jmath_p = \sqrt{1-e_p^2}$. \\
$i_p$					& Inclination of binary $p$. \\
$\omega_p$				& Argument of periapsis of binary $p$. \\
$\Omega_p$				& Longitude of the ascending node of binary $p$. \\
$\ve{r}_\mathrm{per}$		& Position vector of the perturber relative to the barycentre of the internal system. \\
$e_\mathrm{per}$			& Eccentricity of the perturber's hyperbolic orbit ($e_\mathrm{per}>1$). \\
$q_\mathrm{per}$			& Periapsis distance of the perturber's hyperbolic orbit relative to the barycentre of the internal system. \\
$\ve{e}_\mathrm{per}$		& Eccentricity vector of the perturber's orbit. \\
$\ve{\jmath}_\mathrm{per}$	& Normalized angular-momentum vector of the perturber's orbit. \\
$\mathcal{A}_m^{(n)}$		& Two-index dimensionless coefficient appearing in the Legendre polynomials (see equation~\ref{eq:A_mn}). \\
$\mathcal{B}_{i_1,i_2}^{(n,m)} (e_p)$	& Four-index dimensionless function of the eccentricity of binary $p$ (see equation~\ref{eq:B_nm_i1i2}). \\
$\mathcal{D}^{(n,i_1,i_2)}_{l_1,l_2,l_3,l_4}(e_p,e_\mathrm{per})$	& Seven-index dimensionless function of the eccentricities of binary $p$ and of the hyperbolic orbit (see equation~\ref{eq:app:D_def}). \\
$\ve{R}_i$					& Position vector of body $i$ (relative to an arbitrary inertial frame). \\
$\ve{V}_i$					& Velocity vector of body $i$ (relative to an arbitrary inertial frame). \\
$\ve{r}_p$			& Relative position vector of binary $p$. \\
$\ve{v}_p$			& Relative velocity vector of binary $p$. \\
$\ma{A}$					& Mass ratio matrix that relates binary and body coordinates (see equation~\ref{eq:R_to_r}). \\
\bottomrule
\end{tabular}
\caption{Description of important symbols used throughout this paper. }
\label{table:sym}
\end{table}

\begin{figure}
\center
\includegraphics[scale = 0.6, trim = 10mm 20mm 0mm 0mm]{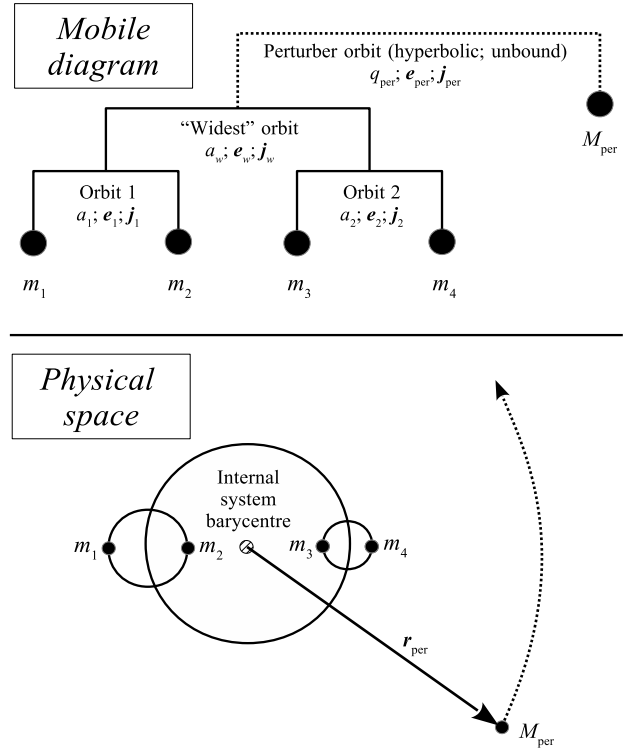}
\caption {Schematic depiction of an example 2+2 quadruple system perturbed by a fifth body on an `external', unbound hyperbolic orbit. Top: a representation of the system in a mobile diagram \citep{1968QJRAS...9..388E}. Bottom: top view of the system in physical space (not to scale), for the simplest case that all orbits in the quadruple system are circular and coplanar, and the perturber is coplanar with respect to the quadruple system. }
\label{fig:cartoon}
\end{figure}

We consider a hierarchical multiple system with an arbitrary number of bodies and structure composed of nested binary orbits (the `internal' system), which is perturbed by a more distant body with mass $M_\mathrm{per}$. An overview with descriptions of the most important symbols used throughout this paper is given in Table~\ref{table:sym}, and a schematic depiction for an example 2+2 quadruple system with an external perturber is given in \F\,\ref{fig:cartoon}. The perturber is assumed to follow a prescribed orbit, $\ve{r}_\mathrm{per} = \ve{r}_\mathrm{per}(t)$, with respect to the barycentre of the internal system. We describe the dynamics using a Hamiltonian formalism. In Appendix \ref{app:per:gen}, we derive the Hamiltonian of the system, which is given by the Hamiltonian associated with the internal system, $H_\mathrm{int}$, plus additional terms associated with the perturbing body, $H_\mathrm{per}$. Expressions for $H_\mathrm{int}$, including its fully orbit-averaged version, were given in \hpz. Note that in the simplest case of a binary perturbed by a passing body, the internal system does not evolve in the absence of the perturber.

If only pairwise interactions between orbits are taken into account, then $H_\mathrm{per}$ consists of terms that are applied to each of the orbits in the internal system. This implies that the perturbations can simply be added individually to each of the orbits in the internal system. In Appendix \ref{app:per:gen}, we show that the non-pairwise terms, i.e., terms that individually depend on three or more orbits, are typically small compared to the pairwise terms, even for systems with equal component masses in the orbits. Note, however, that indirect coupling between the orbits is still possible. For example, in a strongly nested system such as a multiplanet system with widely separated orbits, perturbations from the external body are typically small on the innermost orbits, but larger on the outermost orbits. Subsequently, secular interactions between the orbits of the planets can `transmit' the perturbations to the innermost system. This type of evolution was examined by, e.g., \citet{2004AJ....128..869Z}.

We assume the `secular' or `slow' regime of perturbations, in which the angular frequencies of the orbits in the internal system are much larger than the angular frequency of the perturber. In this regime, the energies of the bound orbits remain conserved, whereas the eccentricities and orbital orientations do not. Quantitatively, for hyperbolic perturbations the secular approximation implies that $\dot{f}_\mathrm{per} \ll \tilde{n}_w$, where $f_\mathrm{per}$ is the true anomaly of the orbit of the perturber, evaluated at periapsis, and $\tilde{n}_w$ is the mean motion of the orbit in the internal system with the longest orbital period\footnote{The mean motion is denoted with `$\tilde{n}$' rather than just `$n$'; `$n$' is reserved below to indicate the Hamiltonian expansion order.}. Equivalently, for hyperbolic perturbations the ratio
\begin{align}
\label{eq:R_def}
\mathcal{R}_w \equiv \frac{\dot{f}_\mathrm{per}}{\tilde{n}_w} = \left [ \left ( 1 + \frac{M_\mathrm{per}}{M_w} \right ) \left (\frac{a_w}{q_\mathrm{per}} \right )^3 \left ( 1 + e_\mathrm{per} \right ) \right ]^{1/2} \ll 1,
\end{align}
where $q_\mathrm{per}$ and $e_\mathrm{per}$ are the periapsis distance and eccentricity of the external hyperbolic orbit, respectively, and $a_w$ is the semimajor axis of orbit $w$, the orbit in the internal system with the longest orbital period. Owing to the definition in terms of the true and mean anomalies, we will refer to the ratio in equation~(\ref{eq:R_def}) as the `angular speed ratio'. Evidently, $\dot{f}_\mathrm{per} \ll \tilde{n}_w$ for distant encounters ($q_\mathrm{per} \gg a_w$) that are not highly eccentric, and with perturber masses not much more massive than the internal system. In this regime, i.e., the secular regime, it is a good approximation to average over the orbits in the internal system, at least as far as perturbations from the external orbit are concerned. 

In addition, we assume that $r_w/r_\mathrm{per}$, where $r_w$ is the orbital separation of the widest binary in the internal system, is sufficiently small at all times compared to $r_\mathrm{per}$ such that it is justified to expand the Hamiltonian in terms of the small parameter $r_w/r_\mathrm{per} \ll 1$. This approximation is also known as the `tidal' approximation. Within a factor of 1 to 2, the largest value of $r_w/r_\mathrm{per}$ is the same as $a_w/q_\mathrm{per}$, showing that secular encounters ($\dot{f}_\mathrm{per} \ll \tilde{n}_w$) are necessarily tidal, but tidal encounters are not necessarily secular. In particular, a tidal encounter may not be secular if $M_\mathrm{per}/M_w$ and/or $e_\mathrm{per}$ are large. We have derived expressions of the expanded Hamiltonian taking into account pairwise interactions only, and valid for arbitrary order $n$, i.e., $H \propto (a_w/r)^n$. In the algorithm implementation, this was restricted to $n\leq5$ for approach (1), and $n\leq 3$ for approach (2) (see \S s \ref{sect:sec:comp:a1} and \ref{sect:sec:comp:a2} below). In our experience, in most practical situations it suffices to include the two lowest-order terms, i.e., terms to the quadrupole ($n=2$) and octupole ($n=3$) order.

In our approximation, the perturber moves on a prescribed orbit with a fixed angular momentum vector $\ve{L}_\mathrm{per}$ (note that with $\ve{L}$ we do {\it not} denote the the angular momentum vector per unit mass, also known as the specific angular momentum vector). Strictly speaking, this implies that the total angular momentum of the system, $\ve{L}_\mathrm{tot} = \ve{L}_\mathrm{int} + \ve{L}_\mathrm{per}$ (internal system + perturber), is not conserved, since in reality the perturber's orbit changes in response to the angular-momentum exchange between the internal system and the perturber, keeping $\ve{L}_\mathrm{tot}$ constant. Our approximation is valid in the limit that the perturber carries much more angular momentum than the internal system, such that perturber's angular momentum is not much affected by the encounter. Since by construction the angular momentum of the internal system is dominated by the widest orbit $w$, this limit applies if $L_w \ll L_\mathrm{per}$, i.e., if
\begin{align}
\label{eq:AM_limit}
\frac{L_w}{L_\mathrm{per}} = \frac{M_{w.\mathrm{C1}} M_{w.\mathrm{C2}}}{M_w M_\mathrm{per}} \left [ \left (1 + \frac{M_\mathrm{per}}{M_w} \right ) \left ( \frac{a_w}{q_\mathrm{per}} \right ) \frac{1-e_w^2}{1+e_\mathrm{per}} \right ]^{1/2} \ll 1.
\end{align}
Here, $M_w$ denotes the total mass of binary $w$, and $M_{w.\mathrm{C}i}$ denotes the mass of all bodies contained within child $i$ of binary $w$. Equation~(\ref{eq:AM_limit}) somewhat resembles equation~(\ref{eq:R_def}); if $q_\mathrm{per} \gg a_w$, then $L_w \ll L_\mathrm{per}$, unless $M_\mathrm{per} \gg M_w$. In other words, secular encounters typically satisfy the condition that the perturber's angular momentum is much larger than the internal system's. 

Note that $\dot{f}_\mathrm{per} \gg \tilde{n}_w$ corresponds to impulsive encounters. In the impulse approximation, the motion of the bodies in the internal system is negligible compared to the fast passage of the perturber, and the latter effectively imparts an impulsive kick to the bodies in the internal system. This process in principle changes the orbital energies, angular momenta and orientations. The impulsive regime is discussed in the context of instantaneous orbital changes, in \S\,\ref{sect:inst:imp}.

\subsection{Computing the orbital changes}
\label{sect:sec:comp}
The secular approximation, $\dot{f}_\mathrm{per} \ll \tilde{n}_w$, implies that the passage time-scale of the perturber is much longer than the orbital period of any of the binaries in the internal system. Therefore, in the secular approximation it is appropriate to average the Hamiltonian over the internal orbits. From the result $H=H_\mathrm{int} + H_\mathrm{per}$ (see Appendix \ref{app:per:gen}), it follows that the orbital vectors of all binaries $p$ in the internal system change according to
\begin{subequations}
\label{eq:EOM_gen}
\begin{align}
\dot{\ve{e}}_p &= \dot{\ve{e}}_{p; \,\mathrm{int}} + \dot{\ve{e}}_{p; \,\mathrm{per}}; \\
\dot{\ve{\jmath}}_p &= \dot{\ve{\jmath}}_{p; \,\mathrm{int}} + \dot{\ve{\jmath}}_{p; \,\mathrm{per}}.
\end{align}
\end{subequations}
Here, $\ve{e}_p$ and $\ve{\jmath}_p$ are the eccentricity and (normalized) angular momenta vectors, respectively, of binary $p$ (with $\jmath_p = \sqrt{1-e_p^2}$). The `internal' terms $\dot{\ve{e}}_{p; \,\mathrm{int}}$ and $\dot{\ve{\jmath}}_{p; \,\mathrm{int}}$ depend on the orbital vectors of other binaries in the internal system, and follow from the equations given in \hpz. For pairwise interactions, the terms associated with the perturber are given by

\begin{subequations}
\label{eq:EOM_r_per}
\begin{align}
\nonumber & \dot{\ve{e}}_{p; \,\mathrm{per}} = \tilde{n}_p \frac{M_\mathrm{per}}{M_p} \sum_{n=2}^\infty (-1)^n \mathcal{M}_p^{(n)} \left ( \frac{a_p}{r_\mathrm{per}} \right )^{n+1} \sum_{m=0}^n \mathcal{A}_m^{(n)} \\
\nonumber &\quad \times  \sum_{\substack{i_1,i_2 \in \, \mathbb{N}^0 \\ i_1+i_2 \leq m}} r_\mathrm{per}^{-i_1-i_2} \Biggl [ i_2 \mathcal{B}_{i_1,i_2}^{(n,m)} (e_p)  \\
\nonumber & \quad \quad  \times \left ( \ve{e}_p \cdot \ve{r}_\mathrm{per} \right )^{i_1} \left ( \ve{\jmath}_p \cdot \ve{r}_\mathrm{per} \right )^{i_2-1} \left ( \ve{e}_p \times \ve{r}_\mathrm{per} \right )  \\
\nonumber &\quad  + \left( \ve{\jmath}_p \cdot \ve{r}_\mathrm{per} \right )^{i_2} \Biggl \{ \frac{\mathrm{d} \mathcal{B}_{i_1,i_2}^{(n,m)}}{\mathrm{d} e_p} \left ( \ve{e}_p \cdot \ve{r}_\mathrm{per} \right )^{i_1} \left ( \ve{\jmath}_p \times \ve{e}_p \right ) \\
&\quad \quad + i_1 \mathcal{B}_{i_1,i_2}^{(n,m)} \left ( \ve{e}_p \cdot \ve{r}_\mathrm{per} \right )^{i_1-1} \left ( \ve{\jmath}_p \times \ve{r}_\mathrm{per} \right ) \Biggl \} \Biggl ]; \\
\nonumber & \dot{\ve{\jmath}}_{p; \,\mathrm{per}} = \tilde{n}_p \frac{M_\mathrm{per}}{M_p} \sum_{n=2}^\infty (-1)^n \mathcal{M}_p^{(n)} \left ( \frac{a_p}{r_\mathrm{per}} \right )^{n+1} \sum_{m=0}^n \mathcal{A}_m^{(n)} \\
\nonumber &\quad \times \sum_{\substack{i_1,i_2 \in \, \mathbb{N}^0 \\ i_1+i_2 \leq m}} r_\mathrm{per}^{-i_1-i_2} \mathcal{B}_{i_1,i_2}^{(n,m)} (e_p) \left [ i_2 \left ( \ve{e}_p \cdot \ve{r}_\mathrm{per} \right )^{i_1} \right. \\
\nonumber & \quad \quad \left. \times  \left ( \ve{\jmath}_p \cdot \ve{r}_\mathrm{per} \right )^{i_2-1} \left ( \ve{\jmath}_p \times \ve{r}_\mathrm{per} \right ) \right. \\
&\quad \left. + i_1 \left ( \ve{e}_p \cdot \ve{r}_\mathrm{per} \right )^{i_1-1} \left ( \ve{\jmath}_p \cdot \ve{r}_\mathrm{per} \right )^{i_2} \left ( \ve{e}_p \times \ve{r}_\mathrm{per} \right ) \right ].
\end{align}
\end{subequations}
These expressions are written in terms of the mean motion of binary $p$, $\tilde{n}_p \equiv \left [GM_p/a_p^3 \right ]^{1/2}$. For notational convenience, we introduced the mass ratio $\mathcal{M}_p^{(n)}$ given by
\begin{align}
\label{eq:mathcal_M}
\mathcal{M}_p^{(n)} \equiv \frac{ \left |  M_{p.\mathrm{C2}}^{n-1} + (-1)^n M_{p.\mathrm{C1}}^{n-1} \right | }{M_p^{n-1}},
\end{align}
where $M_p$ denotes the total mass of binary $p$, and $M_{p.\mathrm{C}i}$ denotes the mass of all bodies contained within child $i$ of binary $p$. The coefficient $\mathcal{A}_m^{(n)}$ and the function $\mathcal{B}_{i_1,i_2}^{(n,m)} (e_p)$ are defined in equations~(\ref{eq:A_mn}) and (\ref{eq:B_nm_i1i2}), respectively. 

Based on equations~(\ref{eq:EOM_r_per}), we proceed with the following approaches for computing the effects of the perturber on the internal system.

\begin{enumerate}[label=\arabic*)]
\item Numerically integrating the equations of motion for the orbital vectors assuming either hyperbolic (1a) or straight-line (1b) orbits $\ve{r}_\mathrm{per}(t)$. The internal orbital vectors change progressively in time as the perturber passes by.
\item Analytically integrating the equations of motion (assuming a hyperbolic perturbation), assuming that the internal orbital vectors do not change during the flyby. 
\end{enumerate}

A hyperbolic orbit represents the correct motion in the limit that the perturber is unbound with respect to the internal system and that its motion is not affected by the quadrupole (and higher) moments of the internal system, i.e., the internal system is effectively a point mass. In the limit of high perturber eccentricity ($e_\mathrm{per}\gg 1$), a hyperbolic orbit can be approximated by a straight line. The different approaches are discussed in more detail below.

\subsubsection{Approach (1)}
\label{sect:sec:comp:a1}
For a binary perturbed by a passing third body, $\dot{\ve{e}}_{p; \,\mathrm{int}} =  \dot{\ve{\jmath}}_{p; \,\mathrm{int}} = \ve{0}$. Nevertheless, even for a straight-line orbit, $\ve{r}_\mathrm{per}(t) = \ve{r}_\mathrm{per,0} + \dot{\ve{r}}_\mathrm{per} t$, and taking into account only the lowest-order terms corresponding to $n=2$, the nonlinear differential equations~(\ref{eq:EOM_gen}) are not amenable to analytic solutions. The same conclusion applies to hyperbolic orbits. Of course, it is possible to numerically integrate equations~(\ref{eq:EOM_r_per}) for a prescribed function $\ve{r}_\mathrm{per}(t)$. This is used in approach (1), for both hyperbolic (1a) and straight-line orbits (1b). Note that a hyperbolic orbit is more general than a straight-line, but numerically also slightly more demanding since it requires numerical solutions to the Kepler equation.

\subsubsection{Approach (2)}
\label{sect:sec:comp:a2}
The right-hand-sides of equations~(\ref{eq:EOM_r_per}) depend on the orbital vectors themselves; the latter change instantaneously as the perturber moves along its specified path (and mostly at periapsis passage). In practice, for weak encounters, the changes of the orbital vectors during the passage are small. Therefore, one can make a further approximation in which the equations of motion are integrated analytically over the passage of the perturber, assuming that the orbital vectors of the internal system, $\ve{e}_p$ and $\ve{\jmath}_p$, do not change. The main advantage of this (strictly inconsistent) approach is that the orbital changes can be computed directly from analytic expressions, without having to numerically integrate the equations of motion. This assumption was also made by \citep{1996MNRAS.282.1064H} for encounters with binaries, with the order of the expansion $n\leq 3$. We derived formal expressions for arbitrary order $n\geq 2$ and assuming pairwise interactions. They are given by
\begin{subequations}
\label{eq:delta_e_n}
\begin{align}
\nonumber \Delta \ve{e}_p &= \frac{\tilde{n}_p}{\tilde{n}_\mathrm{per}} \frac{ M_\mathrm{per}}{M_p} \sum_{n=2}^\infty (-1)^{n} \mathcal{M}_p^{(n)}  \left ( \frac{a_p}{q_\mathrm{per}} \right )^{n+1}  \frac{ \left (e_\mathrm{per}^2-1\right )^{3/2}}{\left (1+e_\mathrm{per}\right)^{n+1}} \\
\nonumber & \times \sum_{m=0}^n \mathcal{A}_m^{(n)} \sum_{\substack{i_1,i_2 \in \, \mathbb{N}^0 \\ i_1+i_2 \leq m}}  \sum_{\substack{l_1,l_2,l_3,l_4 \in \, \mathbb{N}^0 \\ l_1+l_3 \leq i_1 \\ l_2 + l_4\leq i_2}} \\
&\quad \times \ve{f}_{\Delta \ve{e}_p}(\ve{e}_p,\ve{\jmath}_p,\ve{e}_\mathrm{per},\unit{\jmath}_\mathrm{per}; n,m,i_1,i_2,l_1,l_2,l_3,l_4);
\end{align}
\begin{align}
\nonumber \Delta \ve{\jmath}_p &= \frac{\tilde{n}_p}{\tilde{n}_\mathrm{per}} \frac{ M_\mathrm{per}}{M_p}  \sum_{n=2}^\infty (-1)^{n}  \mathcal{M}_p^{(n)}  \left ( \frac{a_p}{q_\mathrm{per}} \right )^{n+1} \frac{ \left (e_\mathrm{per}^2-1\right )^{3/2}}{\left (1+e_\mathrm{per}\right)^{n+1}} \\
\nonumber &\quad \times \sum_{m=0}^n \mathcal{A}_m^{(n)} \sum_{\substack{i_1,i_2 \in \, \mathbb{N}^0 \\ i_1+i_2 \leq m}}  \sum_{\substack{l_1,l_2,l_3,l_4 \in \, \mathbb{N}^0 \\ l_1+l_3 \leq i_1 \\ l_2 + l_4\leq i_2}} \\
&\quad \times \ve{f}_{\Delta \ve{\jmath}_p}(\ve{e}_p,\ve{\jmath}_p,\ve{e}_\mathrm{per},\unit{\jmath}_\mathrm{per}; n,m,i_1,i_2,l_1,l_2,l_3,l_4).
\end{align}
\end{subequations}
Here, $\tilde{n}_\mathrm{per} \equiv \left [G(M_\mathrm{per}+M_\mathrm{int})/|a_\mathrm{per}|^3 \right ]^{1/2}$ is the mean hyperbolic motion, and $|a_\mathrm{per}| = q_\mathrm{per}/(e_\mathrm{per}-1)$. The functions $\ve{f}_{\Delta \ve{e}_p}$ and $\ve{f}_{\Delta \ve{\jmath}_p}$ are defined in terms of the functions $\mathcal{B}_{i_1,i_2}^{(n,m)} (e_p)$ and $\mathcal{D}^{(n,i_1,i_2)}_{l_1,l_2,l_3,l_4}(e_p,e_\mathrm{per})$ in equations~(\ref{eq:f_Delta_ep_jp}). For reference, the coefficient $\mathcal{A}_m^{(n)}$ and the functions $\mathcal{B}_{i_1,i_2}^{(n,m)} (e_p)$ and $ \mathcal{D}^{(n,i_1,i_2)}_{l_1,l_2,l_3,l_4}(e_p,e_\mathrm{per})$ for $n=2$ and $n=3$ and the cases when $\mathcal{D}^{(n,i_1,i_2)}_{l_1,l_2,l_3,l_4}(e_p,e_\mathrm{per}) \neq 0$ are given in Table \ref{table:ABD_n23}.

The lowest-order, or quadrupole-order terms ($n=2$), are typically most important, unless $a_p/q_\mathrm{per}$ is relatively large and/or the orbit $p$ is close to circular, as pointed out by \citeauthor{1996MNRAS.282.1064H} (\citeyear{1996MNRAS.282.1064H}; see also \S\,\ref{sect:sec:ver}). The explicit expressions to this order are given by equations~(\ref{eq:delta_e_j_2}) in Appendix~\ref{app:per:appr:ext}. In the latter appendix, we also show that the eccentricity change implied by equations~(\ref{eq:delta_e_j_2}) is consistent with the corresponding result (for hyperbolic encounters with binaries) by \citet{1996MNRAS.282.1064H}.

\subsection{Verification}
\label{sect:sec:ver}

\begin{figure*}
\center
\includegraphics[scale = 0.5, trim = 10mm 0mm 0mm 0mm]{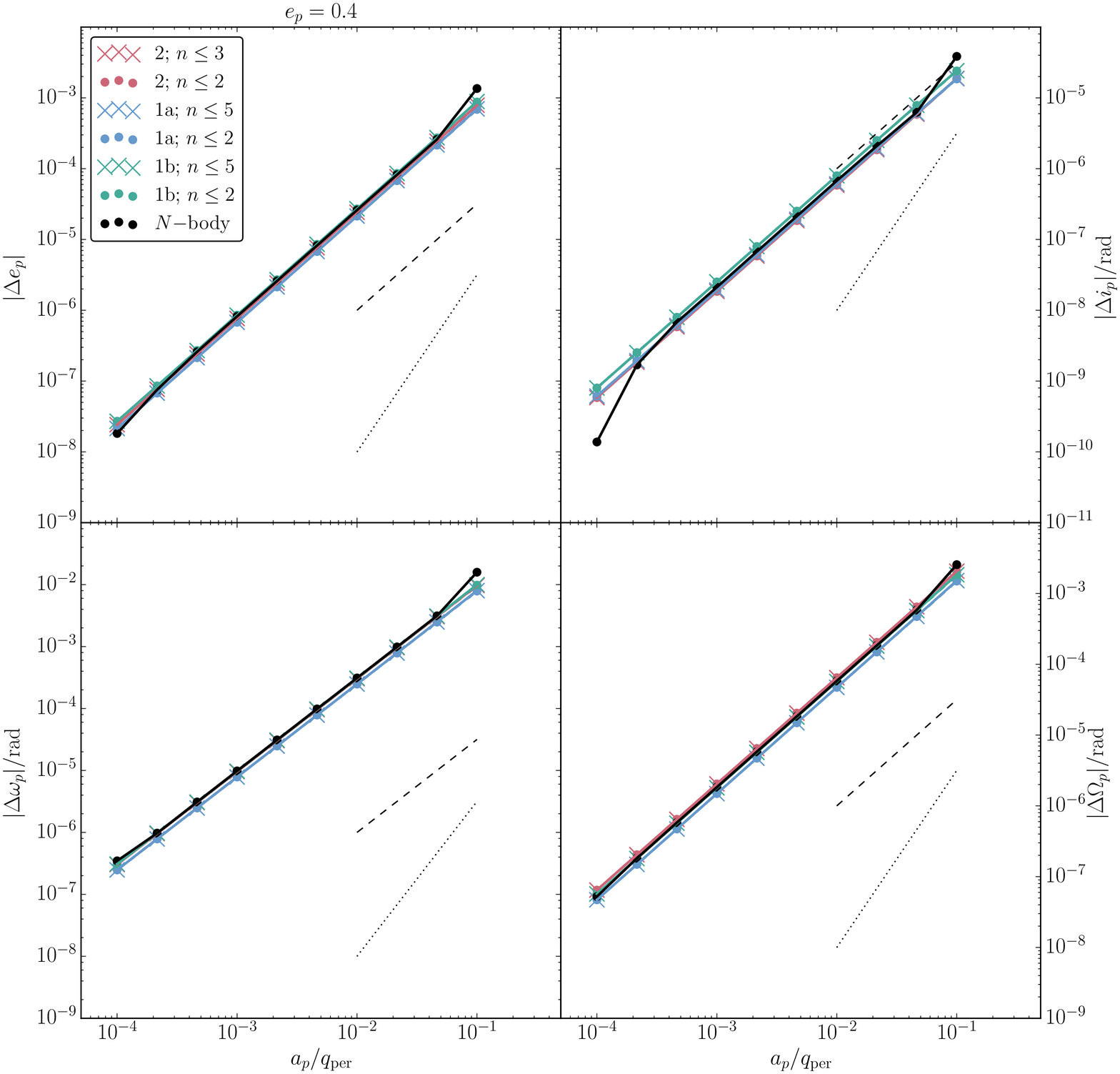}
\caption {Absolute values of the changes of orbital elements of a binary with eccentricity $e_p = 0.4$ after a hyperbolic encounter with a star (discussed in \S\,\ref{sect:sec:ver}). Various binary semimajor axes $a_p$ are considered for a fixed periapsis distance of $q_\mathrm{per}=10^4\,\au$. Shown are the changes of the eccentricity $e_p$, the inclination $i_p$, the argument of periapsis $\omega_p$ and the longitude of the ascending node $\Omega_p$, as a function of $a_p/q_\mathrm{per}$. Different colors correspond to the different secular approaches (see Sections \,\ref{sect:sec:comp:a1} and \,\ref{sect:sec:comp:a2}). For the secular approaches, we include results with the quadrupole-order terms only ($n\leq 2$; crosses) and with terms up and including octupole order ($n \leq 3$) and dotriacontupole order ($n \leq 5$). Solid black lines show results from direct $N$-body integrations. The dashed and dotted lines show a power-law dependence $\propto (a_p/q_\mathrm{per})^{3/2}$ and $\propto (a_p/q_\mathrm{per})^{5/2}$, respectively. }
\label{fig:ver_sma_e_index_e_3_test01}
\end{figure*}

\begin{figure*}
\center
\includegraphics[scale = 0.5, trim = 10mm 0mm 0mm 0mm]{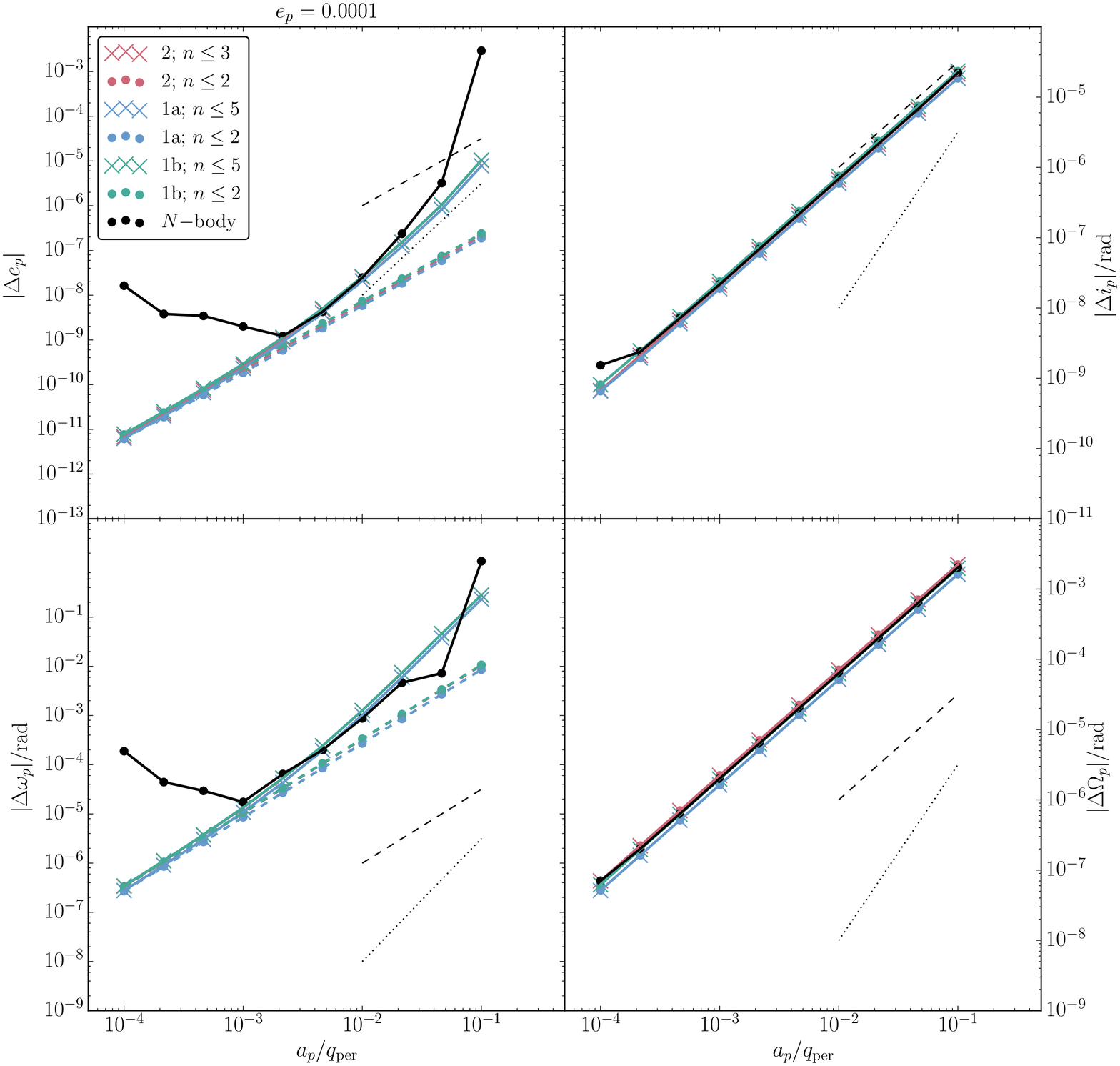}
\caption {Similar to \F\,\ref{fig:ver_sma_e_index_e_3_test01}, now with a nearly circular binary, $e_p = 10^{-4}$. In this case, octupole-order terms are important for large ratios $a_p/q_\mathrm{per}$. }
\label{fig:ver_sma_e_index_e_1_test01}
\end{figure*}

\begin{figure}
\center
\includegraphics[scale = 0.45, trim = 10mm 0mm 0mm 0mm]{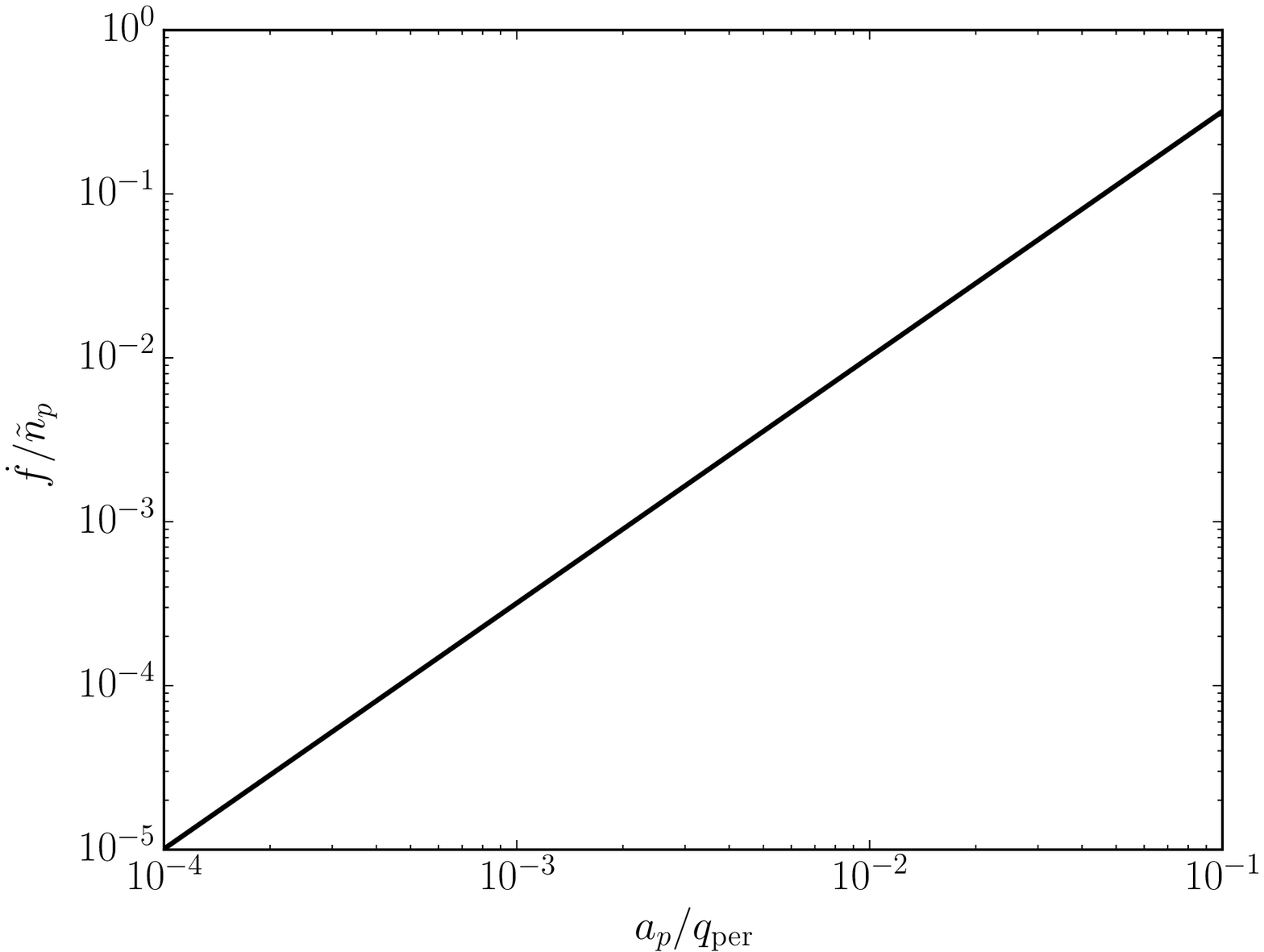}
\caption { The relation between $a_p/q_\mathrm{per}$ and $\dot{f}_\mathrm{per}/\tilde{n}_p$ (see equation~\ref{eq:R_def}) for our chosen parameters in Figs.\,\ref{fig:ver_sma_e_index_e_3_test01} and \ref{fig:ver_sma_e_index_e_1_test01}, i.e., $e_\mathrm{per}\simeq 64.4$ and $M_\mathrm{per}/M_p = 1/1.8$. }
\label{fig:ver_sma_e_ratios_test01}
\end{figure}

We verify our methods by computing the effect of a single flyby on a binary system. We distinguish between the (secular) approaches (1a), (1b) and (2) described in \S\,\ref{sect:sec:comp}, and direct $N$-body integration. As mentioned in \S\,\ref{sect:sec:comp}, implementations (1a) and (1b) are expected to reduce to the same result if $e_\mathrm{per} \gg 1$, in which case the hyperbolic trajectory is effectively a straight line. Furthermore, the assumption made in implementation (2) is, strictly speaking, inconsistent; evidently, in reality, the orbital vectors do change. For small orbital changes, the error made with this inconsistency is expected to be small.

We consider a binary with semimajor axes $a_p$ ranging between $1$ and $10^3$ AU, and various eccentricities $e_p$. The binary component masses are set to 1 and 0.8 $M_\odot$, respectively ($M_p = 1.8\,\msun$); the mass of the perturber is set to $M_\mathrm{per} = 1 \, M_\odot$. The periapsis distance of the perturber orbit is assumed to be $q_\mathrm{per} = 10^4\,\au$ with an eccentricity of $e_\mathrm{per} \simeq 64.4$. The passage is integrated for $t_\mathrm{end} = 0.1\,\mathrm{Myr}$, with periapsis occurring at $t_\mathrm{end}/2$. This is longer than the time-scale associated with periapsis passage, $\pi/\dot{f}_\mathrm{per} \simeq 0.037 \, \mathrm{Myr}$. 

For the $N$-body integrations, we use \textsc{Sakura} \citep{2014MNRAS.440..719G}, a code that uses a Keplerian-based Hamiltonian splitting method to efficiently solve systems with intrinsic hierarchies. Note that accuracy is important for the $N$-body integrations, since the number of binary orbits can be large, up to $\sim 10^5$ for the smallest semimajor axis of 1 AU (for the largest semimajor axis of $10^3\,\au$, the number of orbital revolutions is $\sim 4$). Evidently, large errors in the orbit of the binary can result in spurious changes of the orbital elements. In particular, spurious precession will give incorrect changes of the argument of periapsis, $\Delta \omega_p$. 

In Figs.\,\ref{fig:ver_sma_e_index_e_3_test01} and \ref{fig:ver_sma_e_index_e_1_test01}, we show the changes of the orbital elements of the binary for $e_p = 0.4$ and $e_p=10^{-4}$, respectively. We consider changes of all orbital elements relevant in the secular regime, i.e., $\Delta e_p = \unit{e}_p \cdot \Delta \ve{e}_p$, $\Delta i_p$ (change of inclination), $\Delta \omega_p$ (change of the argument of periapsis) and $\Delta \Omega_p$ (change of the longitude of the ascending node). These changes are plotted as a function of $a_p/q_\mathrm{per}$. For reference, we show in \F\,\ref{fig:ver_sma_e_ratios_test01} the simple relation between $a_p/q_\mathrm{per}$ and $\dot{f}_\mathrm{per}/\tilde{n}_p$ (see equation~\ref{eq:R_def}) for our chosen parameters. We recall that $\dot{f}_\mathrm{per}/\tilde{n}_p \ll 1$ characterizes the secular regime. In the figures, we show results from the secular approximations with the quadrupole-order terms only ($n\leq 2$; crosses) and terms up and including octupole order ($n \leq 3$) and dotriacontupole order ($n \leq 5$).

In the case of a somewhat eccentric binary ($e_p=0.4$; see \F\,\ref{fig:ver_sma_e_index_e_3_test01}), the addition of higher-than-quadrupole-order terms does not affect the results. There is generally good agreement between the different secular approaches, and the direct $N$-body integrations. There are noticeable differences for the smallest and largest ratios $a_p/q_\mathrm{per}$. For the largest ratios $a_p/q_\mathrm{per}$ (and hence $\dot{f}_\mathrm{per}/\tilde{n}_p$), the approximations start to break down. In particular, the `true' eccentricity and orbital orientation changes (i.e., according to the $N$-body integrations) are larger than based on the secular approximations alone; evidently, impulsive-like changes enhance the perturbations. For the smallest ratio, $a_p = 1\,\au$, and the number of orbits is largest. The apparent discrepancy in $\Delta i_p$ between the secular approximations and the direct $N$-body integrations is likely due to errors made by the $N$-body code in the binary motion.

For a nearly circular binary ($e_p=10^{-4}$; see \F\,\ref{fig:ver_sma_e_index_e_1_test01}), the higher-than-quadrupole-order terms are more important (this was pointed out previously by \citealt{1996MNRAS.282.1064H}). The secular changes of the eccentricities and inclinations with $n>2$ show a smooth transition from the quadrupole-order power-law dependence $(a_p/q_\mathrm{per})^{3/2}$ for small $a_p/q_\mathrm{per}$, to the octupole-order power-law dependence $(a_p/q_\mathrm{per})^{5/2}$ (note that the quadrupole and octupole-order results are the same with regard to the inclination and the longitude of the ascending node). For intermediate values of $a_p/q_\mathrm{per}$, this dependence is also produced in the $N$-body integrations. For small $a_p/q_\mathrm{per}$, there is a substantial discrepancy with regard to the eccentricities and arguments of periapsis. This is likely because the changes of these quantities become so small that they are comparable to the precision with which the direct $N$-body code can integrate the binary orbits. For large $a_p/q_\mathrm{per}$, the orbital changes according to the $N$-body integrations are larger than those of the secular approximations, again because of the breakdown of the secular approximations.

\subsection{Code usage}
\label{sect:sec:code}
In Code Fragments~\ref{code:sec:init}, \ref{code:sec:app1} and \ref{code:sec:app2}, we give a minimal working example \textsc{Python} script of how to include encounters in \textsc{\codename} as interfaced in \textsc{AMUSE} for the simplest case of a binary (the example script is easy to extend to more complex systems). Here, we assume that the user has installed \textsc{AMUSE}, and has some familiarity with \textsc{Python} and \textsc{AMUSE}; we refer to \href{http://www.amusecode.org}{http://www.amusecode.org} for detailed installation instructions of \textsc{AMUSE} and examples. 

In the first few lines of Code Fragment \ref{code:sec:init}, the system is set up. Both bodies and binaries are part of the \texttt{particles} set (with length $N_\mathrm{bodies}+N_\mathrm{binaries} = N_\mathrm{bodies} + N_\mathrm{bodies}-1 = 2N_\mathrm{bodies}-1$), and the hierarchy of the system is defined therein with the particle properties \texttt{is\_binary} (a Boolean), and \texttt{child1} and \texttt{child2} (symbolic links within the \texttt{particles} set). Subsequently, the perturbers (in this case, only one) are specified by means of another particle set, named \texttt{external\_particles}, which should be added to the \texttt{external\_particles} of the \textsc{Python} instance of \textsc{\codename}. Generally, the following general properties should be set for each external particle:
\begin{enumerate}
\item \texttt{mass}: the perturber mass, $M_\mathrm{per}$;
\item \texttt{t\_ref}: the time of periapsis passage of the perturber in its orbit with respect to the internal system;
\item \texttt{mode}: the secular approach used (integer value); set to `0' to numerically integrate over the perturber orbit (approach 1), and to `1' to analytically integrate over the perturber orbit (approach 2);
\item \texttt{path}: the type of path assumed for the perturber (integer value); set to `0' for straight-line orbits, and to `1' for hyperbolic orbits; note that \texttt{path=0} is currently not supported if \texttt{mode=1}.
\end{enumerate}
For straight-line orbits, $\ve{r}_\mathrm{per} = \ve{r}_0 + \dot{\ve{r}}_0 \, (t-t_\mathrm{ref})$, the following additional properties should be set:
\begin{enumerate}
\item \texttt{r0\_vec\_x}, \texttt{r0\_vec\_y}, \texttt{r0\_vec\_z}: the $x$, $y$ and $z$ components of $\ve{r}_0$;
\item \texttt{rdot\_vec\_x}, \texttt{rdot\_vec\_y}, \texttt{rdot\_vec\_z}: the $x$, $y$ and $z$ components of $\dot{\ve{r}}_0$.
\end{enumerate}
For hyperbolic orbits (as in the example script), the required properties are:
\begin{enumerate}
\item \texttt{periapse\_distance}: the perturber periapsis distance with respect to the barybarycentre of the internal system, $q_\mathrm{per}$;
\item \texttt{eccentricity}: the eccentricity $e_\mathrm{per}$ of the orbit of the perturber (the relation between eccentricity and speed at infinity, $V_\infty$, is $V_\infty^2 = [e_\mathrm{per}-1] G[M_\mathrm{per} + M_\mathrm{in}]/q_\mathrm{per}$);
\item \texttt{e\_hat\_vec\_x}, \texttt{e\_hat\_vec\_y}, \texttt{e\_hat\_vec\_z}: the $x$, $y$ and $z$ components of the unit eccentricity vector of the hyperbolic orbit;
\item \texttt{h\_hat\_vec\_x}, \texttt{h\_hat\_vec\_y}, \texttt{h\_hat\_vec\_z}: the $x$, $y$ and $z$ components of the unit angular momentum vector of the hyperbolic orbit.
\end{enumerate}

In Code Fragment~\ref{code:sec:app1}, approach (1) is adopted, i.e., the hyperbolic orbit is integrated over numerically, and to compute the perturbation a time loop should be used. In Code Fragment~\ref{code:sec:app2}, the fully analytic approach (2) is used, and the perturbation is taken into account by calling the function \texttt{apply\_external\_perturbation\_assuming\_integrated}
\texttt{\_orbits()} of the \texttt{code} instance of \textsc{\codename}. 

For the parameters chosen in the example scripts, the perturbation on the binary is small; the eccentricity changes by $\simeq 4.7 \times 10^{-5}$, and the two approaches (1) and (2) give the same result within $\simeq 0.08\%$. Note that by default, the $n\leq5$ terms are included for approach (1), and the $n\leq3$ terms for approach (2).

In this example, only one perturber was included; more perturbers can be taken into account by adding more particles to the \texttt{external\_particles} particle set. We emphasize, however, that each encounter is treated independently from each other, i.e., when multiple perturbers are present in \texttt{external\_particles}, their effect on the internal system is computed by adding the contributions of each perturber separately.

\subsection{Application: BH triples in globular clusters}
\label{sect:sec:app}
We briefly discuss an application of the implementation of secular flybys within \textsc{\codename}. In particular, we consider the effect of secular encounters on the evolution of BH triples in globular clusters (GCs). Such triples can form in the cores of GCs through dynamical interactions (e.g., \citealt{1993Natur.364..423S,1993Natur.364..421K,2000ApJ...528L..17P}), and subsequent LK cycles can drive the BHs in the inner binary to merge, producing strong gravitational wave signals detectable by aLIGO and VIRGO (e.g., \citealt{2002ApJ...576..894M,2003ApJ...598..419W,2016ApJ...816...65A}). \citet{2016ApJ...816...65A} modeled the dynamical evolution of BH triples with direct $N$-body integrations, with the initial conditions for the triple systems taken from the Monte-Carlo simulations of \citet{2015ApJ...800....9M}. The rates were found to be $\sim 1\,\mathrm{yr^{-1}}$, and semisecular evolution \citep{2012ApJ...757...27A,2014ApJ...781...45A} was found to be important for these systems once the inner orbit eccentricity is high. 

\citet{2016ApJ...816...65A} treated the triple systems as being isolated, and integrated each system until a strong encounter would disrupt the system, thereby bring a halt to the LK evolution. Weaker (secular) encounters, however, were not taken into account, and such encounters could potentially affect the triple system and the LK evolution. Here, we briefly discuss the effects of secular encounters by simulating secular flybys coupled with LK evolution within \textsc{\codename}. We consider systems that are not in the semisecular regime (not highly eccentric in the inner binary).

\begin{figure}
\center
\includegraphics[scale = 0.44, trim = 0mm 0mm 0mm 0mm]{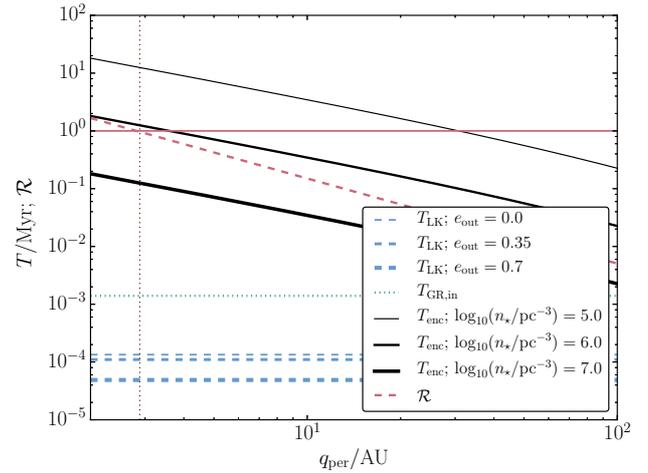}
\caption { Red dashed line: the angular speed ratio $\mathcal{R}$ (defined with respect to the outer binary; see also equation~\ref{eq:R_def}) as a function of $q_\mathrm{per}$. The solid horizontal red line indicates $\mathcal{R}=1$. Solid black lines: the time-scales on which a particular encounter will occur with periapsis distance $q_\mathrm{per}$ (equation~\ref{eq:T_enc}), for different number densities $n_\star$ (refer to the legend). Horizontal blue dashed lines: estimates of the LK time-scales (equation~\ref{eq:T_LK}). Horizontal dotted green line: the relativistic precession time-scale in the inner binary, assuming a circular orbit. }
\label{fig:triple_encounter_timescales.eps}
\end{figure}

We consider a triple with masses $m_1=10\,\msun$, $m_2=8\,\msun$ (inner binary) and $m_3=8\,\msun$ (tertiary). The inner and outer semimajor axes and the (initial) eccentricities are $a_\mathrm{in} = 0.1\,\au$, $a_\mathrm{out} = 2\,\au$, and $e_\mathrm{in}=0.01$, $e_\mathrm{out}=0.7$. The initial mutual inclination is set to $40^\circ$, close to the critical inclination for LK evolution. First, we show for this system in \F\,\ref{fig:triple_encounter_timescales.eps} with the red dashed line the angular speed ratio $\mathcal{R}$ (defined with respect to the outer binary) as a function of $q_\mathrm{per}$. Our interest is in the regime of secular encounters which do not affect the semimajor axes, but can change the eccentricities. The horizontal solid red line shows $\mathcal{R}=1$, which corresponds to $q_\mathrm{per}\simeq 2.9\,\au$. Furthermore, we show in \F\,\ref{fig:triple_encounter_timescales.eps} with solid black lines the time-scales on which a particular encounter will occur with periapsis distance $q_\mathrm{per}$, as a function of $q_\mathrm{per}$. This time-scale can be estimated by \citep[S7.5.8]{2008gady.book.....B}
\begin{align}
\label{eq:T_enc}
\frac{1}{T_\mathrm{enc}} = 4 \sqrt{\pi} n_\star \sigma \left (q_\mathrm{per}^2 + \frac{G(m_1+m_2+m_3+M_\mathrm{per})q_\mathrm{per}}{2 \sigma^2} \right ),
\end{align}
where $n_\star$ is the stellar number density, and $\sigma$ is the one-dimensional velocity dispersion. Here, we adopt a fixed $\sigma=10\,\mathrm{km\,s^{-1}}$, and plot lines corresponding to different densities (refer to the legend in \F\,\ref{fig:triple_encounter_timescales.eps}). For $n_\star = 10^7 \, \mathrm{pc^{-3}}$, $T_\mathrm{enc} \simeq 0.1 \,\mathrm{Myr}$ at $q_\mathrm{per}$ corresponding to $\mathcal{R}=1$. The horizontal blue dashed lines in \F\,\ref{fig:triple_encounter_timescales.eps} show estimates of the LK time-scale (e.g., \citealt{1999CeMDA..75..125K,2015MNRAS.452.3610A}),
\begin{align}
\label{eq:T_LK}
T_\mathrm{LK} \simeq \frac{P_\mathrm{out}^2}{P_\mathrm{in}} \frac{m_1+m_2+m_3}{m_3} \left (1-e_\mathrm{out}^2 \right )^{3/2},
\end{align}
for different values of $e_\mathrm{out}$ (indicated in the legend). Clearly, the LK time-scale is much shorter than the encounter time-scale, indicating that there will be a large number of LK oscillations before the triple is disrupted, or at least its LK evolution is interrupted.

In our simulations, we sample encounters using the same methodology and assumptions of \citet{2017AJ....154..272H}, but reject encounters for which $\mathcal{R}\leq1$ to restrict to secular encounters. Specifically, we assume a locally homogeneous stellar background with stellar number density $n_\star = 10^7 \, \mathrm{pc^{-3}}$ and one-dimensional velocity dispersion $\sigma = 10\,\mathrm{km \, s^{-1}}$, independent of stellar mass. The velocity distribution far away from the system is assumed to be Maxwellian, which is corrected for gravitational focusing due to the triple system. We introduce an `encounter sphere' centered at the barycentre of the triple, with radius $\renc = 100 \, \au \gg a_\mathrm{out}$. Stars impinging on this encounters sphere are considered as perturbers, and their hyperbolic orbit properties with respect to the outer orbit are computed from the perturber mass and velocity. For the perturber mass, we assume a Salpeter distribution \citep{1955ApJ...121..161S}, $\mathrm{d}N/\mathrm{d}M_\mathrm{i}\propto M_\mathrm{i}^{-2.35}$, with lower and upper limits 10 and 100 $\msun$ (the lower limit is set to a value representing a stellar-mass black hole, to model more closely the stellar population within a highly dense GC core, which is expected to be mass-segregated). Using the \textsc{SSE} stellar evolution code \citep{2000MNRAS.315..543H} as implemented in \textsc{AMUSE} \citep{2013CoPhC.183..456P,2013A&A...557A..84P} and assuming a metallicity $Z=0.001$, the initial sampled mass is replaced by the mass after 5 Gyr of stellar evolution. The latter mass is also corrected for gravitational focusing by the triple system. 

We then compute the secular effect of the perturbers within the encounter sphere within \textsc{\codename} using approach (1a) (see \S\,\ref{sect:sec:comp}); in addition to (secular) Newtonian gravity with $n\leq 5$, we include the first post-Newtonian corrections in the inner and outer orbits (the inner orbit relativistic precession time-scale is not much longer than the LK time-scale, see the horizontal green dotted line in \F\,\ref{fig:triple_encounter_timescales.eps}). We check for dynamical stability of the triple in the simulations using the stability criterion of \citet{2001MNRAS.321..398M}, which has recently been found to work well for a wide range of parameters by \citet{2018MNRAS.474...20H}.

\begin{figure*}
\center
\includegraphics[scale = 0.38, trim = 45mm 0mm 0mm 0mm]{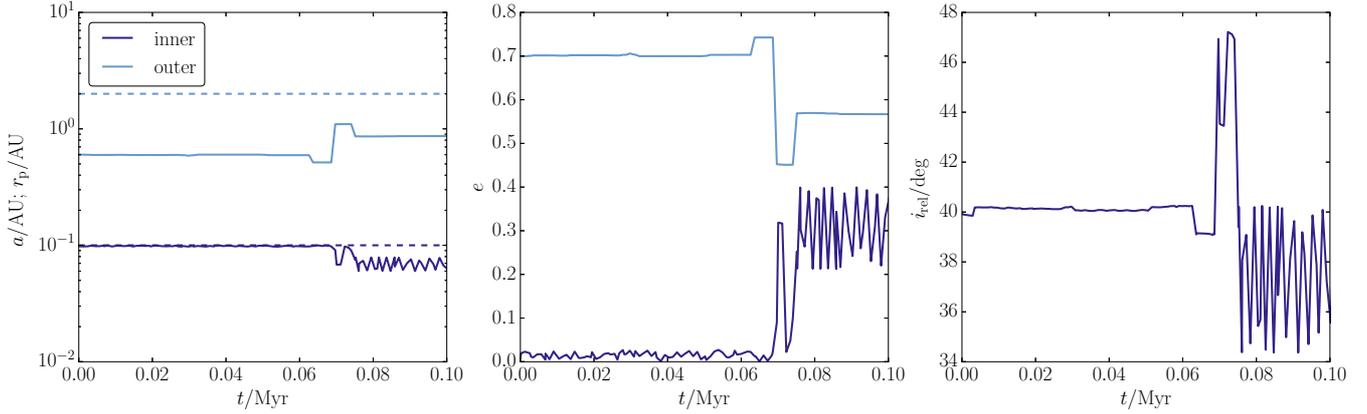}
\caption { Example evolution of a triple in the center of a dense GC subject to external secular flybys, and internal LK evolution. In the left panel, the dashed lines correspond to the semimajor axes, whereas the solid lines correspond to the periapsis distances. Encounters increase the mutual inclination, bringing the triple into a more active regime. See the text in \S\,\ref{sect:sec:app} for details. }
\label{fig:triple_encounter_example}
\end{figure*}

We show an example in \F\,\ref{fig:triple_encounter_example}. In the absence of secular encounters, the triple is not strongly interacting; although the inclination is close to the critical LK angle, the amplitude of the eccentricity oscillations is very small. However, when secular encounters are taken into account, the mutual inclination is increased by several degrees, thereby bringing the triple into a more active regime. Another possibility (not shown here) is that the outer orbit eccentricity is increased, triggering a dynamical instability of the triple. Both these processes can affect the number of merging BH binaries, by either potentially increasing rates in the case of increasing the mutual inclination, or decreasing the rates in the case of destabilizing the triple. A more detailed investigation of these aspects is left for future work.

\section{Instantaneous perturbations}
\label{sect:inst}
In this section, we describe a method to compute the effects of instantaneous perturbations on multiple systems with an arbitrary number of bodies and hierarchy. The method allows for instantaneous changes in the positions, velocities and masses of any of the bodies (not necessarily limited to a single body). In particular, it can be used to compute the effects of SNe with kicks imparted on bodies (in which case the masses and velocities of the bodies change), or impulsive encounters (in which case only the velocities of the bodies change) on these systems. We remark that instantaneous perturbations can change all properties of the hierarchical system, in particular the semimajor axes, and even the hierarchy itself (for example, bodies can be unbound from the system, translating to negative semimajor axes). In contrast, secular encounters (\S\,\ref{sect:sec}) only affect the orbital angular momentum and eccentricity vectors, keeping the semimajor axes (and the hierarchy) fixed. 

First, we give general formulae for the new orbital semimajor axes and eccentricities (\S\,\ref{sect:inst:for}); subsequently, we describe the implementation in \textsc{\codename}, and illustrate the usage of the code (\S\,\ref{sect:inst:impl}). Lastly, we briefly discuss an application to impulsive encounters, and show an example of impulsive encounters with 2+2 quadruple systems (\S\,\ref{sect:inst:imp}).

\subsection{General formulae for orbital changes}
\label{sect:inst:for}
\subsubsection{General case}
\label{sect:inst:for:gen}
Consider a hierarchical $N$-body system composed of nested orbits. Following the same notation as in Appendix A of \hpz, let the mass of body $i$ be denoted with $m_i$, and the position and velocity vectors with respect to an arbitrary origin with $\ve{R}_i$ and $\ve{V}_i$, respectively. We assume that by some process, the masses and the position and velocity vectors change instantaneously according to
\begin{align}
\label{eq:def_changes}
\nonumber m_i' &= m_i + \Delta m_i; \\
\nonumber \ve{R}'_i &= \ve{R}_i + \Delta \ve{R}_i; \\
\ve{V}'_i &= \ve{V}_i + \Delta \ve{V}_i,
\end{align}
where $\Delta m_i$, $\Delta \ve{R}_i$ and $\Delta \ve{V}_i$ are assumed to be known for each body. The specific energy and (squared) angular momentum of each orbit $i$ are given by
\begin{subequations}
\label{eq:E_and_Lsq}
\begin{align}
E_i &= \frac{1}{2} \ve{v}_i^2 - \frac{GM_i}{r_i} = - \frac{GM_i}{2a_i}; \\
h^2_i &= || \ve{r}_i \times \ve{v}_i ||^2 = r_i^2 v_i^2 - \left ( \ve{r}_i \mycdot \ve{v}_i \right )^2 = G M_i a_i \left (1-e_i^2 \right ),
\end{align}
\end{subequations}
where $M_i$ is the mass of all bodies contained within orbit $i$, and $\ve{r}_i$ and $\ve{v}_i$ are the relative separation and velocity vectors, respectively. The change in $M_i$ is 
\begin{align}
\label{eq:Delta_M_i}
\Delta M_i = M_i'-M_i = \sum_{j \in \{i.\mathrm{C} \}} \Delta m_j,
\end{align}
where $j \in \{i.\mathrm{C} \}$ denotes the summation over the bodies in all children of $i$, i.e., $\Delta M_i$ is the sum of all mass changes of the bodies contained within orbit $i$. Let the changes in $\ve{r}_i$ and $\ve{v}_i$ be denoted with $\Delta \ve{r}_i$ and $\Delta \ve{v}_i$, respectively. The latter quantities can be expressed in terms of the mass ratio matrix, $\ma{A}$. As explained in Appendix A of \hpz, the system can be defined in terms of $\ma{A}$ such that the relation between the position vectors of the bodies and the relative separation vectors of the binaries is
\begin{align}
\label{eq:R_to_r}
\ve{r}_i = \sum_{j=1}^N A_{ij} \ve{R}_j,
\end{align}
and similarly for the velocity vectors. The matrix $\ma{A}$ is defined fully in terms of mass ratios of bodies in the system; therefore, equation~(\ref{eq:def_changes}) implies that $\ma{A}$ is affected by the instantaneous perturbation. Let the new matrix, evaluated with the primed masses, be denoted with $\ma{A}'$, and let $\Delta \ma{A} \equiv \ma{A}' - \ma{A}$. The changes in the relative position and velocity vectors are then given by
\begin{subequations}
\label{eq:Delta_r_i_and_v_i}
\begin{align}
\label{eq:Delta_r_i}
\Delta \ve{r}_i &= \sum_{j=1}^N \left [ A_{ij} \Delta \ve{R}_j + \Delta A_{ij} \left ( \ve{R}_j + \Delta \ve{R}_j \right ) \right ]; \\
\label{eq:Delta_v_i}
\Delta \ve{v}_i &= \sum_{j=1}^N \left [ A_{ij} \Delta \ve{V}_j + \Delta A_{ij} \left ( \ve{V}_j + \Delta \ve{V}_j \right ) \right ].
\end{align}
\end{subequations}
Applying equations~(\ref{eq:Delta_M_i}) and (\ref{eq:Delta_r_i_and_v_i}) for the mass and relative position and velocity changes to equations~(\ref{eq:E_and_Lsq}), evaluated prior and after the instantaneous change, we find that the semimajor axis of orbit $i$ is affected according to
\begin{align}
\label{eq:inst_new_a}
\frac{a_i'}{a_i} &= \left (1 +  \frac{\Delta M_i}{M_i} \right ) \left [ 1 - 2 \frac{\ve{v}_i \mycdot \Delta \ve{v}_i}{v_{\mathrm{c},i}^2} - \frac{\Delta \ve{v}_i^2}{v_{\mathrm{c},i}^2} + 2 \frac{a_i}{r_i} \left (1 - \frac{r_i}{r_i'} \frac{M_i+\Delta M_i}{M_i} \right ) \right ]^{-1},
\end{align}
and the eccentricity according to
\begin{align}
\label{eq:inst_new_e}
\nonumber &1-e_i'^2 = \left [ 1 - 2 \frac{\ve{v}_i \mycdot \Delta \ve{v}_i}{v_{\mathrm{c},i}^2} - \frac{\Delta \ve{v}_i^2}{v_{\mathrm{c},i}^2} + 2 \frac{a_i}{r_i} \left (1 - \frac{r_i}{r_i'} \frac{M_i+\Delta M_i}{M_i} \right ) \right ] \\
\nonumber & \times \left ( \frac{M_i}{M_i+\Delta M_i} \right )^2 \Biggl [ 1-e_i^2 + \frac{1}{GM_i a_i} \biggl \{ r_i^2 \left (2 \, \ve{v}_i \mycdot \Delta \ve{v}_i + \Delta \ve{v}_i^2 \right )  \\
\nonumber &   \quad + \left (\ve{v}_i + \Delta \ve{v}_i \right )^2 \left (2 \, \ve{r}_i \mycdot \Delta \ve{r}_i + \Delta \ve{r}_i^2 \right )  \\
\nonumber &   \quad - 2  \left ( \ve{r}_i \mycdot \ve{v}_i \right ) \left (\ve{r}_i \mycdot \Delta \ve{v}_i + \ve{v}_i \mycdot \Delta \ve{r}_i + \Delta \ve{r}_i \mycdot \Delta \ve{v}_i \right )  \\
&   \quad - \left (\ve{r}_i \mycdot \Delta \ve{v}_i + \ve{v}_i \mycdot \Delta \ve{r}_i + \Delta \ve{r}_i \mycdot \Delta \ve{v}_i \right )^2 \biggl \} \Biggl ].
\end{align}
Here,
\begin{align}
v_{\mathrm{c},i}^2 \equiv \frac{GM_i}{a_i}
\end{align}
is the squared circular orbital speed associated with orbit $i$.

\subsubsection{Example: hierarchical triple}
To make the above formulae more concrete, we give an example for a hierarchical triple in which the primary body in the inner binary (labeled body number 1) loses mass by an amount of $-\Delta m_1$ (with $\Delta m_1$ being negative), and receives a velocity kick given by $\ve{V}_\mathrm{k}$. Here, we show that our results are consistent with the equations given in the Appendix of \citet{2016ComAC...3....6T}. For a triple, the matrix $\ma{A}$ can be written as
\begin{align}
\label{eq:A_ex_triple}
\renewcommand\arraystretch{2}
\ma{A} = \begin{pmatrix}
1 & -1 & 0 \\ \frac{m_1}{M_\mathrm{in}} & \frac{m_2}{M_\mathrm{in}} & -1 \\
\frac{m_1}{M} & \frac{m_2}{M} & \frac{m_3}{M}
\end{pmatrix},
\end{align}
where $M_\mathrm{in}\equiv m_1+m_2$ is the inner binary mass, and $M\equiv M_\mathrm{in}+m_3$ is the total system mass. With $\ma{A}$ defined as in equation~(\ref{eq:A_ex_triple}), equation~(\ref{eq:R_to_r}) implies that $\ve{r}_1\equiv \ve{r}_\mathrm{in}$ is the inner orbit separation vector, $\ve{r}_2\equiv \ve{r}_\mathrm{out}$ is the outer orbit separation vector pointing from the barycentre of the inner orbit to the third body, and $\ve{r}_3 \equiv \ve{r}_\mathrm{CM}$ is the centre of mass position vector (similar interpretations apply to the relative velocity vectors $\ve{v}_i$). The new mass ratio matrix is given by
\begin{align}
\label{eq:A_new_ex_triple}
\renewcommand\arraystretch{2}
\ma{A}' = \begin{pmatrix}
1 & -1 & 0 \\
\frac{m_1- \Delta m_1}{M_\mathrm{in}- \Delta m_1} & \frac{m_2}{M_\mathrm{in}- \Delta m_1} & -1 \\
\frac{m_1- \Delta m_1}{M- \Delta m_1} & \frac{m_2}{M- \Delta m_1} & \frac{m_3}{M- \Delta m_1}
\end{pmatrix},
\end{align}
implying that
\begin{align}
\label{eq:Delta_A_ex_triple}
\renewcommand\arraystretch{2}
\Delta \ma{A} = -\Delta m_1 \begin{pmatrix}
0 & 0 & 0 \\ 
\frac{m_2}{(M_\mathrm{in} - \Delta m_1)M_\mathrm{in}} & \frac{-m_2}{(M_\mathrm{in}- \Delta m_1)M_\mathrm{in}} & 0 \\
\frac{m_2+m_3}{M(M-\Delta m_1)} & \frac{-m_2}{M(M-\Delta m_1)} & \frac{-m_3}{M(M-\Delta m_1)} \\
\end{pmatrix}.
\end{align}
Equations~(\ref{eq:Delta_r_i_and_v_i}) then imply
\begin{subequations}
\label{eq:triple_Delta_r_and_v}
\begin{align}
\label{eq:triple_Delta_r_1}
\Delta \ve{r}_1 &= \ve{0}; \\
\label{eq:triple_Delta_r_2}
\Delta \ve{r}_2 &= \frac{-\Delta m_1}{M_\mathrm{in}-\Delta m_1} \frac{m_2}{M_\mathrm{in}} \ve{r}_1; \\
\label{eq:triple_Delta_v_1}
\Delta \ve{v}_1 &= \ve{V}_\mathrm{k}; \\
\label{eq:triple_Delta_v_2}
\Delta \ve{v}_2 &= \frac{-\Delta m_1}{M_\mathrm{in}-\Delta m_1} \left [ \frac{m_2}{M_\mathrm{in}} \ve{v}_1 + \ve{V}_\mathrm{k} \left (1 - \frac{m_1}{\Delta m_1} \right ) \right ].
\end{align}
\end{subequations}
Equations~(\ref{eq:triple_Delta_r_2}) and (\ref{eq:triple_Delta_v_2}) are consistent with equations (57) and (61), respectively, of \citet{2016ComAC...3....6T}. Substituting equations~(\ref{eq:triple_Delta_r_and_v}) into equation~(\ref{eq:inst_new_a}), we find for the new semimajor axes of the inner and outer orbits
\begin{align}
\nonumber \frac{a_1'}{a_1} &= \left (1 - \frac{\Delta m_1}{M_\mathrm{in}} \right ) \left [1 - 2 \frac{ \ve{v}_1 \mycdot \ve{V}_\mathrm{k}}{v_{\mathrm{c},1}^2} - \frac{ \ve{V}_\mathrm{k}^2}{v_{\mathrm{c},1}^2} - 2 \frac{a_1}{r_1} \frac{\Delta m_1}{M_\mathrm{in}} \right ]^{-1}; \\
\frac{a_2'}{a_2} &= \left (1 - \frac{\Delta m_1}{M} \right ) \left [1 - 2 \frac{ \ve{v}_2 \mycdot \Delta \ve{v}_2}{v_{\mathrm{c},2}^2} - \frac{ \Delta \ve{v}_2^2}{v_{\mathrm{c},1}^2} + 2 a_2 \frac{r_2-r_2'}{r_2r_2'}  - 2 \frac{a_2}{r_2'} \frac{\Delta m_1}{M} \right ]^{-1},
\end{align}
which is consistent with equations (63) and (73) of \citet{2016ComAC...3....6T}. Similarly, one can derive the eccentricity changes from equation~(\ref{eq:inst_new_e}) (not given explicitly here), and find that they are consistent with equations (70) and (78) of \citet{2016ComAC...3....6T}\footnote{
Note that there are two typographic errors in equation (78) of \citet{2016ComAC...3....6T}. The correct expression is 
\begin{align}
\nonumber &1-e_\mathrm{out}'^{2} = \left (\frac{m_1+m_2+m_3}{m_1+m_2+m_3-\Delta m_1} \right )^2 \Biggl ( 1 - \frac{2a_\mathrm{out}}{r'_\mathrm{out}}\frac{\Delta m_1}{m_1+m_2+m_3} \\
\nonumber &\quad + 2 a_\mathrm{out} \frac{r_\mathrm{out} - r'_\mathrm{out}}{r_\mathrm{out} r'_\mathrm{out}} - 2 \frac{\left(\ve{v}_\mathrm{out} \cdot \ve{v}_\mathrm{sys} \right )}{v_\mathrm{c,out}^2} - \frac{v_\mathrm{sys}^2}{v_\mathrm{c,out}^2} \Biggl ) \Biggl [ \left(1-e_\mathrm{out}^2\right) \\
\nonumber &\quad + \frac{1}{G(m_1+m_2+m_3)a_\mathrm{out}} \Biggl \{ r_\mathrm{out}^2 \left [2 \left(\ve{v}_\mathrm{out} \cdot \ve{v}_\mathrm{sys} \right ) + v_\mathrm{sys}^2\right ] \\
\nonumber &\quad + \left [-2 \alpha \left(\ve{r}_\mathrm{in}\cdot \ve{r}_\mathrm{out} \right ) + \alpha^2 r_\mathrm{in}^2 \right] \left ( \ve{v}_\mathrm{out} + \ve{v}_\mathrm{sys} \right )^2 + 2 \left(\ve{r}_\mathrm{out} \cdot \ve{v}_\mathrm{out} \right ) \biggl [ \alpha \left(\ve{r}_\mathrm{in}\cdot \ve{v}_\mathrm{out} \right ) \\
\nonumber &\quad - \left(\ve{r}_\mathrm{out} \cdot \ve{v}_\mathrm{sys} \right ) + \alpha \left(\ve{r}_\mathrm{in} \cdot \ve{v}_\mathrm{sys} \right ) \biggl ] - \biggl [ - \alpha \left(\ve{r}_\mathrm{in} \cdot \ve{v}_\mathrm{out} \right ) + \left (\ve{r}_\mathrm{out} \cdot \ve{v}_\mathrm{sys} \right ) \\
\nonumber \quad &- \alpha \left(\ve{r}_\mathrm{in} \cdot \ve{v}_\mathrm{sys} \right ) \biggl ]^2 \Biggl \} \Biggl ].
\end{align}
The algorithm used to compute the effects of SNe in triples in the \textsc{TrES} code of \citet{2016ComAC...3....6T}  does not rely on equation (78) of that paper. Therefore, the typographic errors do not affect the SNe calculations in \textsc{TrES}.
}.

\subsection{Code implementation and usage}
\label{sect:inst:impl}
\subsubsection{Implementation}
In \textsc{\codename}, rather than using equations~(\ref{eq:inst_new_a}) and (\ref{eq:inst_new_e}) directly, we adopt a slightly different and more general approach that allows us to calculate the new orbital orientations in addition to the scalar energy and angular-momentum changes. First, from the eccentricity vector, $\ve{e}_i$, the angular momentum vector, $\ve{h}_i$, and the mean anomaly, we compute the relative position and velocity vectors of each orbit $i$. Subsequently, the initial positions and velocities of all bodies are calculated according to equation~(\ref{eq:R_to_r}), and the instantaneous changes are applied as described by equation~(\ref{eq:def_changes}). The masses of all orbits, $M_i$, are then updated, and the new eccentricity and angular momentum vectors are calculated. From the latter vectors, the new orbital elements are inferred, including the orbital angles. We note that with this method, analytic expressions like equations~(\ref{eq:inst_new_a}) and (\ref{eq:inst_new_e}) could also be derived for changes in the orbital orientations (in particular, the inclinations). Here, however, we instead calculate such changes numerically within \textsc{\codename}. 

The above is implemented within a new function of the \textsc{\codename} code, called \textsc{apply\_user\_specified\_instantaneous\_perturbation()}. This function can be used independently of the part of the code that models the {\it secular} evolution of the system.

\subsubsection{Usage}
\label{sect:inst:impl:use}
We briefly illustrate the usage of the algorithm to compute instantaneous orbital changes within \textsc{\codename}. An example in which one of the stars of a hierarchical sextuple system undergoes an asymmetric SN is given in Code Fragment~\ref{code:inst}. The system consists of six bodies in the configuration of a 2+2 quadruple orbited by a binary. First, the orbital elements are specified. The orbital phases are necessary to compute the original positions and velocities of all bodies. They can be supplied directly to each of the binaries through the property \texttt{true\_anomaly}; alternatively, if for any binary the Boolean parameter \texttt{sample\_orbital\_phases\_randomly} is set to \texttt{True} (by default \texttt{False}), then the orbital phase for that orbit is sampled randomly (i.e., the mean anomaly is assumed to have a flat distribution). The seed of the random number generator used can be set with the code parameter \texttt{orbital\_phases\_random\_seed} (e.g., with the line \texttt{code.parameters.orbital\_phases\_random\_seed = 1} to set the seed to \texttt{1}). 

The mass changes of the bodies (in this case, the body corresponding to \texttt{particles[0]}) are specified by setting the \texttt{instantaneous\_perturbation\_delta\_mass} attribute, and the velocity kicks in the $x$-direction are specified with \texttt{instantaneous\_perturbation\_delta\_velocity\_x}, and similarly for the $y$ and $z$-directions. Subsequently, channels are set up, and the orbital changes are computed by calling the function \texttt{apply\_user\_specified\_instantaneous\_perturbation()} of the \textsc{\codename} \textsc{Python} instance. After a required channel copy, the new orbital elements can be retrieved from the \texttt{particles} set through the \texttt{binaries} subset.

\subsection{Application to impulsive encounters}
\label{sect:inst:imp}
\subsubsection{Formulae}
\label{sect:inst:imp:for}
Impulsive encounters can be considered as a type of instantaneous orbital perturbations specified by equations~(\ref{eq:def_changes}). The methodology described above can be used to include impulsive perturbations within \textsc{\codename}, and here we briefly discuss this application. In the impulsive limit, the position vectors of all bodies are assumed to be constant during the encounter, i.e., $\Delta \ve{R}_i = \ve{0}$ (furthermore, $\Delta m_i=0$). We assume that the perturber is unaffected by the multiple system and that no net force acts on it; therefore, it moves on a straight line with respect to the other bodies with constant velocity, i.e., the position vector of the perturber is given by
\begin{align}
\label{eq:straight_line}
\ve{R}_\mathrm{per}(t) = \ve{b} + \ve{V}_\mathrm{per} \, t,
\end{align}
where $\ve{b}$ is the impact parameter vector, and $\ve{V}_\mathrm{per}$ is the perturber's velocity. The perturber then imparts a velocity kick $\Delta \ve{V}_i$ on each body $i$ given by integrating the acceleration on body $i$, i.e.,
\begin{align}
\label{eq:imp}
\nonumber \Delta \ve{V}_i &= \int_{-\infty}^\infty \mathrm{d} t \, G M_\mathrm{per} \frac{\ve{b} + \ve{V}_\mathrm{per} \, t - \ve{R}_i}{ \left [ \left (\ve{b} - \ve{R}_i \right )^2 + 2\left(\ve{b}-\ve{R}_i \right )\mycdot \ve{V}_\mathrm{per} \, t + \ve{V}_\mathrm{per}^2 \, t^2 \right ]^{3/2}} \\
&= 2 \frac{GM_\mathrm{per}}{V_\mathrm{per}} \frac{\unit{b}_i}{b_i},
\end{align}
where we defined the impact parameter vector with respect to body $i$,
\begin{align}
\ve{b}_i \equiv \ve{b}-\ve{R}_i - \unit{V}_\mathrm{per} \left [ \left (\ve{b}-\ve{R}_i \right ) \mycdot \unit{V}_\mathrm{per} \right ].
\end{align}
Equation~(\ref{eq:imp}), combined with the module discussed in \S\,\ref{sect:inst:impl}, can be used to compute the effect of impulsive perturbations on hierarchical multiple systems. 

\subsubsection{Code usage}
\label{sect:inst:imp:code}
We illustrate the use of equation~(\ref{eq:imp}) for impulsive encounters within \textsc{\codename} with Code Fragment \ref{code:inst_imp} for a hierarchical quadruple system. 

\subsubsection{Example: impulsive encounters with 2+2 quadruple systems}
\label{sect:inst:imp:ex}
We briefly demonstrate the cumulative effect of impulsive encounters with a wide 2+2 quadruple system in the Solar neighborhood. The quadruple system has two inner binaries `A' and `B' with semimajor axes $a_\mathrm{A} = 100\,\au$ and $a_\mathrm{B} = 200\,\au$; the superorbit `C' has a semimajor axis of $a_\mathrm{C}=2\times10^4\,\au$. The masses in the `A' subsystem are $m_1=1 \,\msun$ and $m_2=0.8\,\msun$; for the `B' we set $m_3=1\,\msun$ and $m_4=0.9\,\msun$. The initial eccentricities are assumed to be $e_\mathrm{A} = 0.01$, $e_\mathrm{B} = 0.01$ and $e_\mathrm{C}=0.7$, the individual inclinations are $i_\mathrm{A}=55^\circ$, $i_\mathrm{B}=40^\circ$ and $i_\mathrm{C}=0^\circ$, the arguments of periapsis are $\omega_\mathrm{A}=\omega_\mathrm{B}=\omega_\mathrm{C}=0^\circ$, and the longitudes of the ascending nodes are $\Omega_\mathrm{A}=\Omega_\mathrm{B}=\Omega_\mathrm{C}=0^\circ$. These parameters imply initial mutual inclinations of the binaries relative to their parent orbits of $i_\mathrm{AC} = 55^\circ$ and $i_\mathrm{BC} = 40^\circ$ (also, $i_\mathrm{AB}=15^\circ$). The (initial) LK time-scales for the AB and AC pairs (see equation 7 of \hpz) are of the same order of magnitude, i.e., $T_\mathrm{LK,AB} \simeq 2.1\,\mathrm{Gyr}$ and $T_\mathrm{LK,BC} \simeq 0.79\,\mathrm{Gyr}$, respectively, indicating that there could be some weak coupling between the orbits \citep{2017MNRAS.470.1657H}.

We sample encounters using a similar methodology as in \citet{2017AJ....154..272H} and in \S\,\ref{sect:sec:app}, modified by assuming straight-line orbits for the perturbers (see equation~\ref{eq:straight_line}). In particular, from the sampled encounter sphere vector $\ve{R}_\mathrm{enc}$ and the velocity $V_\mathrm{per}$ of the perturber relative to the barycentre of the multiple system at the encounter sphere, we compute the impact parameter according to $\ve{b} = \ve{R}_\mathrm{enc} - \left ( \unit{V}_\mathrm{per} \mycdot \ve{R}_\mathrm{enc} \right ) \unit{V}_\mathrm{per}$. We set the lowest mass of the perturbers to $0.1\,\msun$, and assume a stellar number density of $n_\star = 0.1 \, \mathrm{pc^{-3}}$ and a one-dimensional velocity dispersion of $\sigma=40\,\mathrm{km\,s^{-1}}$ to model encounters in the Solar neighborhood. The perturber mass function is assumed to be a Salpeter function \citep{1955ApJ...121..161S} corrected for gravitational focusing, the stellar age in the initial-final mass relation is assumed to be $10\,\mathrm{Gyr}$, and we assume a metallicity of $Z=0.01$. The encounter sphere radius is set to $R_\mathrm{enc}=1\times 10^4\,\au < a_\mathrm{C}$, such that all sampled encounters with respect to the C orbit are in the impulsive regime (here, we neglect secular and `intermediate' encounters). The impulsive encounters are treated as instantaneous perturbations within \textsc{\codename}; in between encounters, the system is treated as isolated and only the secular interactions are modeled. 

\begin{figure*}
\center
\includegraphics[scale = 0.38, trim = 45mm 0mm 0mm 0mm]{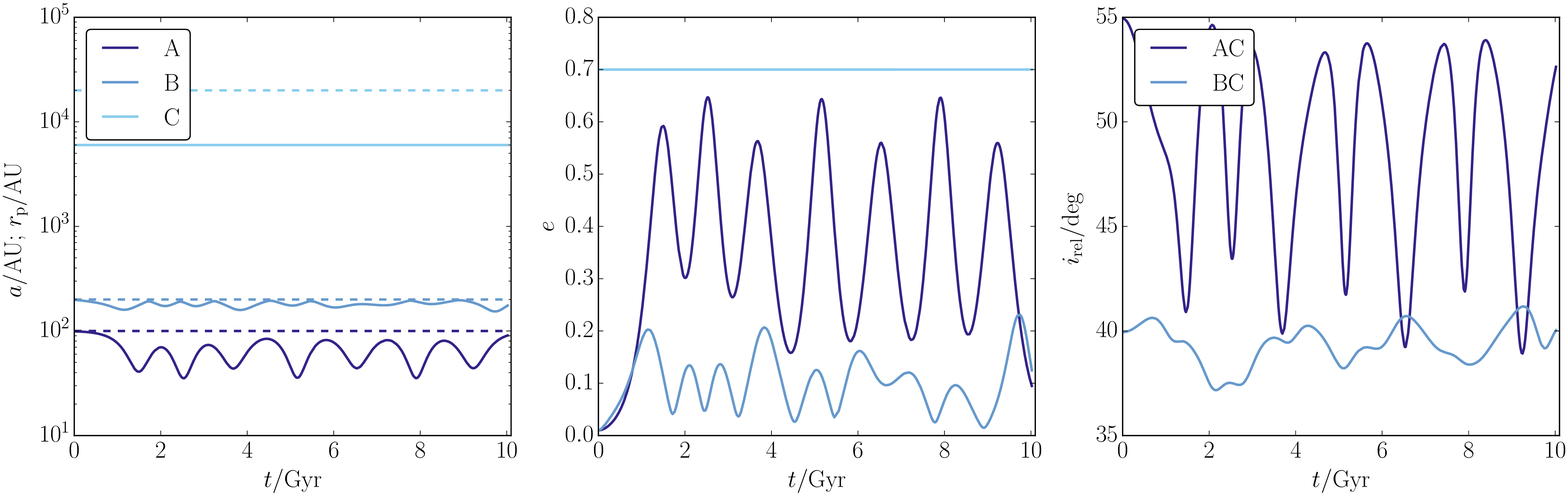}
\includegraphics[scale = 0.38, trim = 45mm 0mm 0mm 0mm]{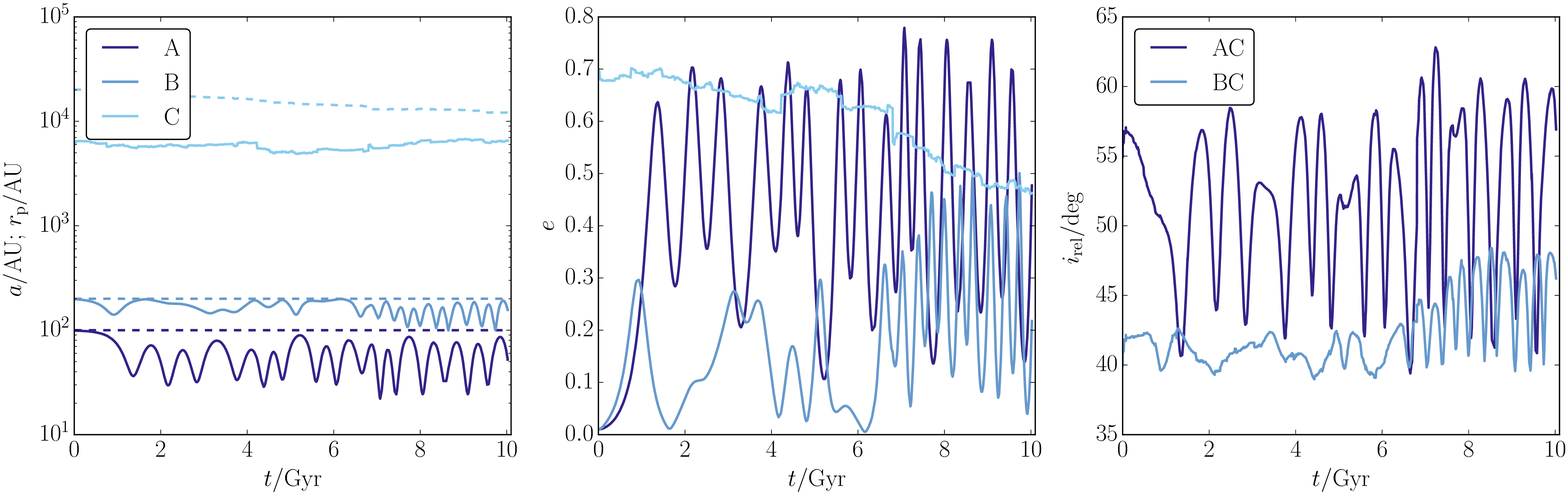}
\caption { Example evolution of a 2+2 quadruple experiencing impulsive flybys. In the top panels, the system is evolved in isolation, whereas in the bottom panels, the system is embedded in the Solar neighborhood. In the left panels, the dashed lines correspond to the semimajor axes, whereas the solid lines correspond to the periapsis distances. Refer to the text in \S\,\ref{sect:inst:imp:ex} for details. }
\label{fig:ex_imp}
\end{figure*}

We show results of one realization of the simulations in \F\,\ref{fig:ex_imp}. In the top panels, the quadruple system is evolved in isolation (i.e., encounters were not taken into account), whereas in the bottom panels, the system was embedded in the Solar neighborhood. In the absence of encounters, the eccentricity of `A' reaches a value close to the expected value if `B' is replaced by a point mass (assuming the test-particle limit), i.e., $\left [1-(5/3) \cos(55^\circ) \right ]^{1/2} \simeq 0.67$. For orbit `B', the uncoupled expected maximum eccentricity is $\left [1-(5/3) \cos(40^\circ) \right ]^{1/2} \simeq 0.15$; the actual value attained, $\approx 0.2$, is slightly larger, likely due to weak coupling between the orbits. 

When encounters are included, orbit C is significantly affected in terms of its semimajor axis, eccentricity and orientation. The A and B orbits now attain higher eccentricities; in particular, $e_\mathrm{A}$ approaches 0.8, whereas $e_\mathrm{B}$ reaches 0.5, which can be attributed to the encounter-induced mutual inclination between orbits B and C of up to $\approx 48^\circ$ (corresponding to $\left [1-(5/3) \cos(48^\circ) \right ]^{1/2} \simeq 0.50$). 

A more detailed investigation into the impact of encounters on the evolution of 2+2 quadruple systems is left for future work.

\section{Discussion}
\label{sect:discussion}

\subsection{Limitations of the encounter treatments}
\label{sect:discussion:enc}
We discussed flybys in both the secular (\S\,\ref{sect:sec}) and impulsive (\S\,\ref{sect:inst:imp}) regimes. Depending on the situation, most encounters might be of one particular type, and one can apply the results of this paper to model both regimes. However, the intermediate regime, $\mathcal{R}\sim 1$, is not well described by either approximations, and this should be taken into account when using the \textsc{\codename} code to treat encounters. To model the intermediate regime, one would have to resort to direct $N$-body integrations; however, this is straightforward to implement since \textsc{\codename} is part of \textsc{AMUSE}, which has a large range of direct $N$-body integrators available, and the positions and velocities of all the bodies in the system can be easily obtained (see, e.g., Code Fragment \ref{code:inst_imp}).

In addition, our formalism only allows for encounters with single stars, and not with higher-order multiple systems. Multiple-multiple encounters are beyond the scope of this work, but they could nonetheless be important. For example, \citet{2015MNRAS.448..344L} showed that binaries are substantially more efficient at perturbing planetary systems compared to single stars due to their larger effective cross section.

\subsection{Breakdown of the orbit-average approximation}
\label{sect:discussion:sec}
As in \hpz, we averaged over all orbits in the internal system, whereas there are situations in which the averaging approach breaks down. In particular, there can be evection-type resonances (e.g., \citealt{2014MNRAS.439.1079A,2016MNRAS.458.3060L}), or, in the case of high eccentricities, the effective LK time-scale can be shorter than the orbital time-scales \citep{2012ApJ...757...27A,2012arXiv1211.4584K}. When using the \textsc{\codename} code, one should be aware of these limitations.

\section{Conclusions}
\label{sect:conclusions}
We expanded upon \hpz\, by adding `external' perturbations to the `internal' hierarchical multiple system, which can be composed of an arbitrary number of bodies and structure, provided it consists of nested binary orbits. The external perturbations included are stellar encounters in the secular regime, and instantaneous orbital changes such as asymmetric SNe and impulsive encounters. Our main conclusions are listed below.

\medskip \noindent 1. For encounters in the secular regime, we showed that the Hamiltonian can be composed of an `internal' part associated with the bound orbits (excluding the perturber), and an `external' part associated with the perturber. Non-pairwise terms (i.e., terms depending on three or more orbits simultaneously) are typically not important. If non-pairwise interactions are ignored, then the external Hamiltonian is obtained by summing pairwise terms associated with the perturber over all binaries in the internal system. Following a similar approach as in \citet{1996MNRAS.282.1064H}, we expanded the external part of the Hamiltonian in terms of ratios of the separations of orbits in the internal system to the perturber, which are assumed to be small. To compute the effects on the internal system, we averaged over the internal orbits in the general case; if the orbital changes are small, then it is a good approximation to also analytically integrate over the perturber orbit, as in \citet{1996MNRAS.282.1064H}. 

\medskip \noindent 2. We derived general expressions for the pairwise terms in the equations of motion for secular encounters at any order, and tested our algorithm with direct $N$-body integrations. We illustrated the usage of the algorithm within \textsc{\codename} in \textsc{AMUSE}, and gave a brief example of secular encounters with BH triples in GCs.

\medskip \noindent 3. We presented a method to include the effects of instantaneous perturbations on the bodies in the multiple system. The method is general; any of the bodies' properties can be changed (mass, position and velocity). It can be used to compute the effects of asymmetric SNe and impulsive encounters, and is included within \textsc{\codename}. We showed how to apply the method to SNe and impulsive encounters in high-order multiple systems, and gave an example of a wide 2+2 quadruple system perturbed by encounters in the Solar neighborhood. 

The extensions presented in this paper are a next step toward efficiently modeling the evolution of complex multiple systems embedded in star clusters.

\section*{Acknowledgements}
I thank Scott Tremaine for discussions and comments on an early version of the manuscript, and the anonymous referee for a very helpful report. I gratefully acknowledge support from the Institute for Advanced Study, The Peter Svennilson Membership, and NASA grant NNX14AM24G.

\bibliographystyle{mnras}
\bibliography{literature}

\appendix

\onecolumn

\section{External secular encounters with hierarchical multiple systems}
\label{app:per}
\subsection{General Hamiltonian}
\label{app:per:gen}
Consider a hierarchical multiple system consisting of $N$ bodies arranged in binary orbits. This system is referenced to as the `internal' system, with the set of `internal' binaries denoted with $\mathrm{B}_\mathrm{int}$. The simplest case of an internal system is a binary, but the method also allows for more complicated systems such as hierarchical triple and quadruple systems. We model the dynamics of this internal system using the formalism presented in \hpz, where an expansion was made in terms of ratios of binary separations, which were assumed to be small. 

Perturbations to the internal system are taken into account by adding to the internal system an additional perturbing body on an `external' orbit. The additional body is modeled as a point mass with mass $M_\mathrm{per}$. Its orbit with respect to the internal system, $\ve{r}_\mathrm{per}(t)$, is assumed to be known a priori. If the internal system were approximated by a point mass, this orbit would generally be parabolic or hyperbolic. We make the tidal approximation, i.e., we assume that the distance $r_\mathrm{per}$ of the perturber to the barycentre of the internal system is, at all times, large compared to the largest binary separation in the internal system (the latter was labelled `$r_m$' in \hpz). This justifies an expansion of the Hamiltonian of the system in terms of the small ratios $r_p/r_\mathrm{per}$, where $r_p$ is any of the binary separations in the internal system. 

Following similar derivations as in Appendix A of \hpz \,(in particular, see equation A101), the expanded (but unaveraged) Hamiltonian of the system is given by
\begin{align}
\label{eq:app:H}
H = H_\mathrm{int} + H_\mathrm{ext},
\end{align}
where the Hamiltonian associated with the internal system is given by\footnote{Note that the factor $1/2$ in equation A101 of \hpz\, was unfortunately misplaced; it should only multiply the kinetic terms. }
\begin{align}
H_\mathrm{int} &= \sum_{k\in\mathrm{B_{int}}} \left [ \frac{1}{2} \frac{  M_{k.\mathrm{C1}} M_{k.\mathrm{C2}}}{M_k} \left (\dot{\ve{r}}_k \cdot \dot{\ve{r}}_k\right )
- \frac{GM_{k.\mathrm{C1}} M_{k.\mathrm{C2}}}{r_k} \right ] + \sum_{n=2}^{\infty} S'_{n; \, \mathrm{int}}.
\end{align}
Here, $n$ is the expansion order of the binary separation ratios, $M_{k.\mathrm{C}i}$ denotes the mass of all bodies contained within child $i$ of binary $k$, and $\ve{r}_k$ is the separation vector of binary $k$. The disturbing function $S'_{n; \, \mathrm{int}}$ was given in Equation (A97) of \hpz, with the binary $k$ now running over all internal binaries $\mathrm{B}_\mathrm{int}$. The dominant terms in $S'_n$ are typically the terms associated with pairwise binary interactions, i.e., the terms (see equation A99 of \hpz)
\begin{align}
\nonumber S'_{n;2; \, \mathrm{int}} &= (-1)^{n+1} \sum_{k \in \mathrm{B_{int}}} \sum_{\substack{p \in \mathrm{B_{int}} \\ p \in \{ k.\mathrm{C} \} }} \frac{M_{p.\mathrm{C1}} M_{p.\mathrm{C2}}}{M_p} \alpha(p.\mathrm{C1},k.\mathrm{C2};p)^n \frac{ M_{p.\mathrm{C2}}^{n-1} + (-1)^n M_{p.\mathrm{C1}}^{n-1}}{M_p^{n-1}} \frac{GM_{k.\mathrm{CS}(p)}}{r_k} \sum_{m=0}^n \mathcal{A}_m^{(n)} \frac{ \left ( \ve{r}_p \cdot \ve{r}_k \right )^m r_p^{n-m} }{r_k^{n+m}} .
\end{align}
Here, $\alpha(p.\mathrm{C1},k.\mathrm{C2};p) = \pm 1$ was defined in Appendix A1 of \hpz. The quantity $\mathcal{A}^{(n)}_m$, an integer ratio, is the same as the coefficients appearing in the Legendre polynomials, which can be obtained from Rodrigues's formula, i.e.,
\begin{align}
\label{eq:A_mn}
\sum_{m=0}^n \mathcal{A}_m^{(n)} x^m = \frac{1}{2^n n!} \frac{\mathrm{d}^n }{\mathrm{d} x^n} \left [ \left (x^2 -1 \right )^n \right ].
\end{align}

The Hamiltonian associated with the external system is given by
\begin{align}
H_\mathrm{ext} &= 
\left [ \frac{1}{2}  \frac{  M_\mathrm{int} M_\mathrm{per} }{M_\mathrm{int} + M_\mathrm{per}} \left (\dot{\ve{r}}_\mathrm{per} \cdot \dot{\ve{r}} _\mathrm{per}\right )
- \frac{G M_\mathrm{int} M_\mathrm{per} }{r_\mathrm{per}} \right ] + \sum_{n=2}^{\infty} S'_{n; \, \mathrm{ext}},
\end{align}
where  $\ve{r}_\mathrm{per}$ is the vector pointing from the barycentre of the internal system to the external body, and the total internal mass is
\begin{align}
M_\mathrm{int} \equiv \sum_{i \in \mathrm{B_{int}}} m_i,
\end{align}
i.e., the total mass of all bodies contained within the internal system. The pairwise terms in the external part of the perturbing potential $S'_{n; \, \mathrm{ext}}$ are given by
\begin{align}
\label{eq:S_n}
S'_{n;2; \, \mathrm{ext}} &= (-1)^{n+1} \sum_{\substack{p \in \mathrm{B_{int}} }} \frac{M_{p.\mathrm{C1}} M_{p.\mathrm{C2}}}{M_p} \alpha(p.\mathrm{C1},\mathrm{ext};p)^n \frac{ M_{p.\mathrm{C2}}^{n-1} + (-1)^n M_{p.\mathrm{C1}}^{n-1}}{M_p^{n-1}} \frac{GM_\mathrm{per}}{r_\mathrm{per}} \sum_{m=0}^n \mathcal{A}_m^{(n)} \frac{ \left ( \ve{r}_p \cdot \ve{r}_\mathrm{per} \right )^m r_p^{n-m} }{r_\mathrm{per}^{n+m}}.
\end{align}

The unaveraged Hamiltonian equation~(\ref{eq:app:H}) is exact in the limit that the expansion is taken to infinite order and that all interactions are taken into account (pairwise, triple-wise, etc.). Concerning the external part, starting at the octupole order ($n=3$), triple-wise terms appear (i.e., terms that depend on three orbits simultaneously), provided that the internal system at least contains two orbits (i.e., at least a triple). The triple-wise terms at the octupole-order are explicitly given by
\begin{align}
\label{eq:S_3_3}
\nonumber S'_{3;3; \, \mathrm{ext}} &= \frac{3}{2} \sum_{\substack{p \in \mathrm{B}_\mathrm{int}}} \sum_{\substack{u \in \mathrm{B}_\mathrm{int} \\ p \in \{u.\mathrm{C} \} }}  \frac{M_{p.\mathrm{C1}} M_{p.\mathrm{C2}}}{M_p} \alpha(p,\mathrm{ext}.\mathrm{CS}(p);\mathrm{ext}) \frac{G M_\mathrm{per}}{r_\mathrm{per}} \frac{\alpha(p,\mathrm{ext}.\mathrm{CS}(p);u) M_{u.\mathrm{CS}(p)}}{M_u} \left ( \frac{r_p}{r_\mathrm{per}} \right )^2 \left ( \frac{r_u}{r_\mathrm{per}} \right ) \\
&\quad \quad \times \left [ 5 \left ( \unit{r}_p \cdot \unit{r}_\mathrm{ext} \right )^2 \left ( \unit{r}_u \cdot \unit{r}_\mathrm{per} \right ) - 2 \left ( \unit{r}_p \cdot \unit{r}_\mathrm{per} \right ) \left ( \unit{r}_p \cdot \unit{r}_u \right ) - \left ( \unit{r}_u \cdot \unit{r}_\mathrm{per} \right ) \right ].
\end{align}
We estimate the importance of the triple-wise terms as follows. Consider a triple system, the simplest system in which the triple-wise terms in the external part of the Hamiltonian appear, with separation ratios $x = r_\mathrm{in}/r_\mathrm{out}$, and let $y = r_\mathrm{out}/r_\mathrm{per}$. Assume that the system is highly hierarchical such that $x\ll 1$ and $y \ll 1$, and assume $x\sim y$. From equation~(\ref{eq:S_3_3}), $S'_{3;3;\,\mathrm{ext}} \propto (r_\mathrm{in}/r_\mathrm{per})^2 (r_\mathrm{out}/r_\mathrm{per}) = x^2 y^3$, whereas from equation~(\ref{eq:S_n}), the dominant terms are $S'_{3;2;\,\mathrm{ext}} \propto x^3$ and $S'_{3;2;\,\mathrm{ext}} \propto y^3$, which are $\gg x^2 y^3$, i.e., $S'_{3;3;\,\mathrm{ext}} \ll S'_{3;2;\,\mathrm{ext}}$. This argument does not apply if the masses in the inner binary are equal ($M_{p.\mathrm{C1}} = M_{p.\mathrm{C2}}$), such that the pairwise octupole-order terms vanish. Nonetheless, in that case, $S'_{4;2} \propto x^4$ and $S'_{4;2} \propto y^4$, which are still $\gg x^2 y^3$, i.e., the pairwise hexadecapole-order terms, which do not vanish for equal masses in the inner binary, still dominate over the triple-wise octupole-order terms. 

In subsequent sections, we make several approximations by restricting to pairwise interactions and averaging over some or all of the orbits, in order to gain more analytically tractable expressions and to speed up numerical calculations.

\subsection{Approximations}
\label{app:per:appr}
\subsubsection{Averaging over orbits in the internal system}
\label{app:per:appr:int}
First, we restrict to pairwise interactions and average the Hamiltonian over all orbits in the internal system. The internal Hamiltonian, apart from constant binding energies, is then given by equation (A136) of \hpz. The external part is given by 
\begin{align}
\label{eq:S_n_2_ext_single_av}
 \overline{S'}_{n;2; \, \mathrm{ext}} &= (-1)^{n+1} \sum_{\substack{p \in \mathrm{B_{int}} }} \frac{M_{p.\mathrm{C1}} M_{p.\mathrm{C2}}}{M_p} \mathcal{M}_p^{(n)} \frac{GM_\mathrm{per}}{r_\mathrm{per}} \sum_{m=0}^n \mathcal{A}_m^{(n)} \left ( \frac{a_p}{r_\mathrm{per}} \right)^n  \sum_{\substack{i_1,i_2 \in \, \mathbb{N}^0 \\ i_1+i_2 \leq m}}  \mathcal{B}_{i_1,i_2}^{(n,m)} (e_p) \, r_\mathrm{per}^{-i_1-i_2} \left ( \ve{e}_p \cdot \ve{r}_\mathrm{per} \right )^{i_1} \left ( \ve{\jmath}_p \cdot \ve{r}_\mathrm{per} \right )^{i_2},
\end{align}
where 
\begin{align}
\mathcal{M}_p^{(n)} \equiv \frac{ \left |  M_{p.\mathrm{C2}}^{n-1} + (-1)^n M_{p.\mathrm{C1}}^{n-1} \right | }{M_p^{n-1}},
\end{align}
and the function $\mathcal{B}_{i_1,i_2}^{(n,m)}$ is defined implicitly by (see also section A5.3 of \hpz)
\begin{align}
\label{eq:B_nm_i1i2}
 &\sum_{\substack{i_1,i_2 \in \, \mathbb{N}^0 \\ i_1+i_2 \leq m}} r_\mathrm{per}^{m-i_1-i_2} \left ( \ve{e}_p \cdot \ve{r}_\mathrm{per} \right )^{i_1} \left ( \ve{\jmath}_p \cdot \ve{r}_\mathrm{per} \right )^{i_2} \mathcal{B}_{i_1,i_2}^{(n,m)} (e_p) =  \left ( \frac{1}{2} \right )^{n+1} \frac{1}{(n+1)!}
\frac{ \mathrm{d}^{n+1}}{\mathrm{d} z^{n+1}} \left [  \left ( -e_p z^2 +2z - e_p  \right )^{n-m+1}  \left ( \frac{\widetilde{C}_1}{e_p} z^2 + \widetilde{C}_3 z + \frac{\widetilde{C}_2}{e_p} \right )^m \right ]_{z=0}.
\end{align}
Here, we defined the complex coefficients
\begin{subequations}
\label{eq:tildeC_def}
\begin{align}
\widetilde{C}_1 &\equiv  \left ( \ve{e}_p \cdot \ve{r}_\mathrm{per} \right ) - \complexi \sqrt{1-e_p^2} \, (\pm1 ) \left [ e_p^2 r_\mathrm{per}^2 - \left ( \ve{e}_p \cdot \ve{r}_\mathrm{per} \right )^2 - \frac{e_p^2}{1-e_p^2} \left ( \ve{\jmath}_p \cdot \ve{r}_\mathrm{per} \right)^2 \right ]^{1/2}; \\
\widetilde{C}_2 &\equiv \left ( \ve{e}_p \cdot \ve{r}_\mathrm{per} \right ) + \complexi \sqrt{1-e_p^2} \, (\pm1 ) \left [ e_p^2 r_\mathrm{per}^2 - \left ( \ve{e}_p \cdot \ve{r}_\mathrm{per} \right )^2 - \frac{e_p^2}{1-e_p^2} \left ( \ve{\jmath}_p \cdot \ve{r}_\mathrm{per} \right)^2 \right ]^{1/2}; \\
\widetilde{C}_3 &\equiv -2 \left ( \ve{e}_p \cdot \ve{r}_\mathrm{per} \right ),
\end{align}
\end{subequations}
where $\complexi = \sqrt{-1}$. 

The equations of motion for the orbital vectors follow from the Milankovitch equations \citep{milankovitch_39},
\begin{subequations}
\label{eq:EOM}
\begin{align}
\label{eq:EOM:j}
\frac{\mathrm{d} \boldsymbol{j}_p}{\mathrm{d} t} &= - \frac{1}{\Lambda_p} \left [ \, \boldsymbol{j}_p \times \nabla_{\boldsymbol{j}_p} \overline{H} + \boldsymbol{e}_p \times \nabla_{\boldsymbol{e}_p} \overline{H} \, \right ]; \\
\label{eq:EOM:e}
\frac{\mathrm{d} \boldsymbol{e}_p}{\mathrm{d} t} &= - \frac{1}{\Lambda_p} \left [ \, \boldsymbol{e}_p \times \nabla_{\boldsymbol{j}_p} \overline{H} + \boldsymbol{j}_p \times \nabla_{\boldsymbol{e}_p} \overline{H} \, \right ].
\end{align}
\end{subequations}
Here, $\Lambda_p$ is the angular momentum of orbit $p$ if it were circular,
\begin{align}
\label{eq:Lambda_main}
\Lambda_p = M_{p.\mathrm{C1}} M_{p.\mathrm{C2}} \, \sqrt{ \frac{G a_p}{M_p}}.
\end{align}

From $H = H_\mathrm{int} + H_\mathrm{ext}$, it immediately follows that the equations of motion for the orbital vectors due to the external perturbation can simply be added to those of the internal system, i.e.,
\begin{subequations}
\begin{align}
\dot{\ve{e}}_p &= \dot{\ve{e}}_{p; \,\mathrm{int}} + \dot{\ve{e}}_{p; \,\mathrm{per}}; \\
\dot{\ve{\jmath}}_p &= \dot{\ve{\jmath}}_{p; \,\mathrm{int}} + \dot{\ve{\jmath}}_{p; \,\mathrm{per}},
\end{align}
\end{subequations}
where $\dot{\ve{e}}_{p; \,\mathrm{int}}$ and $\dot{\ve{\jmath}}_{p; \,\mathrm{int}}$ follow from the internal Hamiltonian. Using equation~(\ref{eq:S_n_2_ext_single_av}), the equations of motion due to the external perturbation then read (again including only pairwise interactions)
\begin{subequations}
\label{app:eq:EOM_ext_single_av}
\begin{align}
\nonumber & \dot{\ve{e}}_{p; \,\mathrm{per}} = \tilde{n}_p \frac{M_\mathrm{per}}{M_p} \sum_{n=2}^\infty (-1)^n \mathcal{M}_p^{(n)} \left ( \frac{a_p}{r_\mathrm{per}} \right )^{n+1} \sum_{m=0}^n \mathcal{A}_m^{(n)} \sum_{\substack{i_1,i_2 \in \, \mathbb{N}^0 \\ i_1+i_2 \leq m}} r_\mathrm{per}^{-i_1-i_2} \Biggl [ i_2 \mathcal{B}_{i_1,i_2}^{(n,m)} (e_p) \left ( \ve{e}_p \cdot \ve{r}_\mathrm{per} \right )^{i_1} \left ( \ve{\jmath}_p \cdot \ve{r}_\mathrm{per} \right )^{i_2-1} \left ( \ve{e}_p \times \ve{r}_\mathrm{per} \right ) \\
&\quad + \left( \ve{\jmath}_p \cdot \ve{r}_\mathrm{per} \right )^{i_2} \Biggl \{ \frac{\mathrm{d} \mathcal{B}_{i_1,i_2}^{(n,m)}}{\mathrm{d} e_p} \left ( \ve{e}_p \cdot \ve{r}_\mathrm{per} \right )^{i_1} \left ( \ve{\jmath}_p \times \ve{e}_p \right ) + i_1 \mathcal{B}_{i_1,i_2}^{(n,m)} \left ( \ve{e}_p \cdot \ve{r}_\mathrm{per} \right )^{i_1-1} \left ( \ve{\jmath}_p \times \ve{r}_\mathrm{per} \right ) \Biggl \} \Biggl ]; \\
\nonumber & \dot{\ve{\jmath}}_{p; \,\mathrm{per}} = \tilde{n}_p \frac{M_\mathrm{per}}{M_p} \sum_{n=2}^\infty (-1)^n \mathcal{M}_p^{(n)} \left ( \frac{a_p}{r_\mathrm{per}} \right )^{n+1} \sum_{m=0}^n \mathcal{A}_m^{(n)}\sum_{\substack{i_1,i_2 \in \, \mathbb{N}^0 \\ i_1+i_2 \leq m}} r_\mathrm{per}^{-i_1-i_2} \mathcal{B}_{i_1,i_2}^{(n,m)} (e_p) \left [ i_2 \left ( \ve{e}_p \cdot \ve{r}_\mathrm{per} \right )^{i_1} \left ( \ve{\jmath}_p \cdot \ve{r}_\mathrm{per} \right )^{i_2-1} \left ( \ve{\jmath}_p \times \ve{r}_\mathrm{per} \right ) \right. \\
&\quad \left. + i_1 \left ( \ve{e}_p \cdot \ve{r}_\mathrm{per} \right )^{i_1-1} \left ( \ve{\jmath}_p \cdot \ve{r}_\mathrm{per} \right )^{i_2} \left ( \ve{e}_p \times \ve{r}_\mathrm{per} \right ) \right ].
\end{align}
\end{subequations}
Here, $\tilde{n}_p \equiv [ G M_p/a_p^3]^{1/2}$ is the mean motion of orbit $p$.

\subsubsection{Integrating over the external orbit}
\label{app:per:appr:ext}
In this approximation, with the equations already averaged over the internal orbits, we analytically integrate the equations of motion~(\ref{app:eq:EOM_ext_single_av}) over time (from $-\infty$ to $\infty$), assuming that the external body moves in a hyperbolic orbit with respect to the barycentre of the internal system. Also, we assume that the changes of the orbital vectors $\ve{e}_p$ and $\ve{\jmath}_p$ during the encounter are small, such that they can be taken to be constant on the right-hand-sides of equation~(\ref{app:eq:EOM_ext_single_av}). A major advantage of this approach, which was also adopted by \citet{1996MNRAS.282.1064H}, is speed: instead of having to numerically integrate over the orbit, the changes of the orbital vectors over the full encounter can be computed directly from analytic expressions, without having to numerically resolve the encounter. Of course, the approach is valid only in the limit that the changes of the orbital vectors are small.

Restricting to pairwise interactions, we find the following expressions for the time-integrated equations of motion, i.e., the changes of the orbital vectors.
\begin{subequations}
\label{app:eq:delta_e_n}
\begin{align}
\Delta \ve{e}_p &= \frac{\tilde{n}_p}{\tilde{n}_\mathrm{per}} \frac{ M_\mathrm{per}}{M_p} \sum_{n=2}^\infty (-1)^{n} \mathcal{M}_p^{(n)}  \left ( \frac{a_p}{q_\mathrm{per}} \right )^{n+1}  \frac{ \left (e_\mathrm{per}^2-1\right )^{3/2}}{\left (1+e_\mathrm{per}\right)^{n+1}} \sum_{m=0}^n \mathcal{A}_m^{(n)} \sum_{\substack{i_1,i_2 \in \, \mathbb{N}^0 \\ i_1+i_2 \leq m}}  \sum_{\substack{l_1,l_2,l_3,l_4 \in \, \mathbb{N}^0 \\ l_1+l_3 \leq i_1 \\ l_2 + l_4\leq i_2}} \ve{f}_{\Delta \ve{e}_p}(\ve{e}_p,\ve{\jmath}_p,\ve{e}_\mathrm{per},\unit{\jmath}_\mathrm{per}; n,m,i_1,i_2,l_1,l_2,l_3,l_4); \\
\Delta \ve{\jmath}_p &= \frac{\tilde{n}_p}{\tilde{n}_\mathrm{per}} \frac{ M_\mathrm{per}}{M_p}  \sum_{n=2}^\infty (-1)^{n}  \mathcal{M}_p^{(n)}  \left ( \frac{a_p}{q_\mathrm{per}} \right )^{n+1} \frac{ \left (e_\mathrm{per}^2-1\right )^{3/2}}{\left (1+e_\mathrm{per}\right)^{n+1}} \sum_{m=0}^n \mathcal{A}_m^{(n)} \sum_{\substack{i_1,i_2 \in \, \mathbb{N}^0 \\ i_1+i_2 \leq m}}  \sum_{\substack{l_1,l_2,l_3,l_4 \in \, \mathbb{N}^0 \\ l_1+l_3 \leq i_1 \\ l_2 + l_4\leq i_2}} \ve{f}_{\Delta \ve{\jmath}_p}(\ve{e}_p,\ve{\jmath}_p,\ve{e}_\mathrm{per},\unit{\jmath}_\mathrm{per}; n,m,i_1,i_2,l_1,l_2,l_3,l_4).
\end{align}
\end{subequations}

Here, $\tilde{n}_\mathrm{per} \equiv \left [G(M_\mathrm{per}+M_\mathrm{int})/|a_\mathrm{per}|^3 \right ]^{1/2}$ is the mean hyperbolic motion with $|a_\mathrm{per}| = q_\mathrm{per}/(e_\mathrm{per}-1)$. The functions $\ve{f}_{\Delta \ve{e}_p}$ and $\ve{f}_{\Delta \ve{\jmath}_p}$ are defined in terms of $\mathcal{B}_{i_1,i_2}^{(n,m)}$ (see equation~\ref{eq:B_nm_i1i2}) and the function $\mathcal{D}^{(n,i_1,i_2)}_{l_1,l_2,l_3,l_4}(e_p,e_\mathrm{per})$; the latter is defined implicitly via
\begin{align}
\label{eq:app:D_def}
\nonumber &\sum_{\substack{l_1,l_2,l_3,l_4 \in \, \mathbb{N}^0 \\ l_1+l_3 \leq i_1 \\ l_2 + l_4\leq i_2}}  \mathcal{D}^{(n,i_1,i_2)}_{l_1,l_2,l_3,l_4}(e_p,e_\mathrm{per}) \left( \ve{e}_p \cdot \unit{e}_\mathrm{per} \right )^{l_1} \left( \ve{\jmath}_p \cdot \unit{\jmath}_\mathrm{per} \right )^{l_2} \left( \ve{e}_p \cdot \unit{\jmath}_\mathrm{per} \right )^{l_3} \left( \ve{\jmath}_p \cdot \unit{e}_\mathrm{per} \right )^{l_4} \equiv \int_{-\mathrm{arccos}(-1/e_\mathrm{per})}^{\mathrm{arccos}(-1/e_\mathrm{per})} \mathrm{d} f \left [1 + e_\mathrm{per} \cos(f) \right ]^{n-1} \\
&\quad \times \left [ \cos(f) \left ( \ve{e}_p\cdot \unit{e}_\mathrm{per} \right ) + \sin(f) \left ( \ve{e}_p\cdot \unit{q}_\mathrm{per} \right ) \right ]^{i_1} \left [ \cos(f) \left ( \ve{\jmath}_p\cdot \unit{e}_\mathrm{per} \right ) + \sin(f) \left ( \ve{\jmath}_p\cdot \unit{q}_\mathrm{per} \right ) \right ]^{i_2},
\end{align}
where $\unit{q}_\mathrm{per} \equiv \unit{\jmath}_\mathrm{per} \times \unit{e}_\mathrm{per}$. To derive $\mathcal{D}^{(n,i_1,i_2)}_{l_1,l_2,l_3,l_4}(e_p,e_\mathrm{per})$ from equation~(\ref{eq:app:D_def}), the following vector identity can be used,
\begin{align}
\label{eq:vec_id}
\left ( \unit{q}_\mathrm{per} \cdot \boldsymbol{u} \right ) \left ( \unit{q}_\mathrm{per} \cdot \boldsymbol{w} \right ) &= \left [ \left ( \unit{\jmath}_\mathrm{per} \times \unit{e}_\mathrm{per} \right ) \cdot \boldsymbol{u} \right ] \left [ \left ( \unit{\jmath}_\mathrm{per} \times \unit{e}_\mathrm{per} \right ) \cdot \boldsymbol{w} \right ] = \boldsymbol{u} \cdot \boldsymbol{w} - \left ( \unit{e}_\mathrm{per} \cdot \boldsymbol{u} \right ) \left ( \unit{e}_\mathrm{per} \cdot \boldsymbol{w} \right ) - \left ( \unit{\jmath}_\mathrm{per} \cdot \boldsymbol{u} \right ) \left ( \unit{\jmath}_\mathrm{per} \cdot \boldsymbol{w} \right ),
\end{align}
where $\ve{u}$ and $\ve{w}$ are arbitrary vectors. The functions $\ve{f}_{\Delta \ve{e}_p}$ and $\ve{f}_{\Delta \ve{\jmath}_p}$ read
\begin{subequations}
\label{eq:f_Delta_ep_jp}
\begin{align}
\nonumber &\ve{f}_{\Delta \ve{e}_p}(\ve{e}_p,\ve{\jmath}_p,\ve{e}_\mathrm{per},\unit{\jmath}_\mathrm{per}; n,m,i_1,i_2,l_1,l_2,l_3,l_4) \equiv \mathcal{B}_{i_1,i_2}^{(n,m)} (e_p) \mathcal{D}^{(n,i_1,i_2)}_{l_1,l_2,l_3,l_4}(e_p,e_\mathrm{per}) \left [ \left ( \ve{e}_p\cdot \unit{e}_\mathrm{per} \right )^{l_1} \left ( \ve{e}_p\cdot \unit{\jmath}_\mathrm{per} \right )^{l_3} \right. \\
\nonumber &\quad \times \left. \left \{ l_2 \left ( \ve{\jmath}_p\cdot \unit{\jmath}_\mathrm{per} \right )^{l_2-1} \left ( \ve{\jmath}_p \cdot \unit{e}_\mathrm{per} \right )^{l_4} \left ( \ve{e}_p \times \unit{\jmath}_\mathrm{per} \right ) + l_4 \left ( \ve{\jmath}_p\cdot \unit{\jmath}_\mathrm{per} \right )^{l_2} \left ( \ve{\jmath}_p \cdot \unit{e}_\mathrm{per} \right )^{l_4-1} \left ( \ve{e}_p \times \unit{e}_\mathrm{per} \right ) \right \} 
+ \left ( \ve{\jmath}_p\cdot \unit{\jmath}_\mathrm{per} \right )^{l_2} \left ( \ve{\jmath}_p \cdot \unit{e}_\mathrm{per} \right )^{l_4} \right. \\
\nonumber &\quad \left. \times \left \{ l_1 \left ( \ve{e}_p\cdot \unit{e}_\mathrm{per} \right )^{l_1-1} \left ( \ve{e}_p\cdot \unit{\jmath}_\mathrm{per} \right )^{l_3}  \left(\ve{\jmath}_p \times \unit{e}_\mathrm{per} \right ) + l_3 \left ( \ve{e}_p\cdot \unit{e}_\mathrm{per} \right )^{l_1}  \left ( \ve{e}_p\cdot \unit{\jmath}_\mathrm{per} \right )^{l_3-1} \left ( \ve{\jmath}_p \times \unit{\jmath}_\mathrm{per} \right ) \right \} \right ] \\
\nonumber &+ \left ( \ve{e}_p\cdot \unit{e}_\mathrm{per} \right )^{l_1} \left ( \ve{\jmath}_p\cdot \unit{\jmath}_\mathrm{per} \right )^{l_2} \left ( \ve{e}_p\cdot \unit{\jmath}_\mathrm{per} \right )^{l_3}  \left ( \ve{\jmath}_p \cdot \unit{e}_\mathrm{per} \right )^{l_4} \\
&\quad \times \left [ \frac{\partial \mathcal{B}_{i_1,i_2}^{(n,m)} (e_p)}{\partial e_p} \mathcal{D}^{(n,i_1,i_2)}_{l_1,l_2,l_3,l_4}(e_p,e_\mathrm{per}) + \mathcal{B}_{i_1,i_2}^{(n,m)} (e_p) \frac{\partial \mathcal{D}^{(n,i_1,i_2)}_{l_1,l_2,l_3,l_4}(e_p,e_\mathrm{per})}{\partial e_p} \right ] \left ( \ve{\jmath}_p \times \unit{e}_p \right ); \\
\nonumber &\ve{f}_{\Delta \ve{\jmath}_p}(\ve{e}_p,\ve{\jmath}_p,\ve{e},\unit{\jmath}_\mathrm{per}; n,m,i_1,i_2,l_1,l_2,l_3,l_4) \equiv \mathcal{B}_{i_1,i_2}^{(n,m)} (e_p) \mathcal{D}^{(n,i_1,i_2)}_{l_1,l_2,l_3,l_4}(e_p,e_\mathrm{per}) \left [ \left ( \ve{e}_p\cdot \unit{e}_\mathrm{per} \right )^{l_1} \left ( \ve{e}_p\cdot \unit{\jmath}_\mathrm{per} \right )^{l_3} \right. \\
\nonumber &\quad \left. \times \left \{ l_2 \left ( \ve{\jmath}_p\cdot \unit{\jmath}_\mathrm{per} \right )^{l_2-1} \left ( \ve{\jmath}_p\cdot \unit{e}_\mathrm{per} \right )^{l_4} \left ( \ve{\jmath}_p \times \unit{\jmath}_\mathrm{per} \right ) + l_4 \left ( \ve{\jmath}_p\cdot \unit{\jmath}_\mathrm{per} \right )^{l_2} \left ( \ve{\jmath}_p\cdot \unit{e}_\mathrm{per} \right )^{l_4-1} \left ( \ve{\jmath}_p \times \unit{e}_\mathrm{per} \right ) \right \} 
+ \left ( \ve{\jmath}_p\cdot \unit{\jmath}_\mathrm{per} \right )^{l_2} \left ( \ve{\jmath}_p \cdot \unit{e}_\mathrm{per}  \right )^{l_4} \right. \\
&\quad \left. \times \left \{ l_1 \left ( \ve{e}_p\cdot \unit{e}_\mathrm{per} \right )^{l_1-1} \left ( \ve{e}_p\cdot \unit{\jmath}_\mathrm{per} \right )^{l_3}  \left(\ve{e}_p \times \unit{e}_\mathrm{per} \right ) + l_3 \left ( \ve{e}_p\cdot \unit{e}_\mathrm{per} \right )^{l_1}  \left ( \ve{e}_p\cdot \unit{\jmath}_\mathrm{per} \right )^{l_3-1} \left ( \ve{e}_p \times \unit{\jmath}_\mathrm{per} \right ) \right \} \right ].
\end{align}
\end{subequations}

For reference, the coefficient $\mathcal{A}_m^{(n)}$ and the functions $\mathcal{B}_{i_1,i_2}^{(n,m)} (e_p)$ and $ \mathcal{D}^{(n,i_1,i_2)}_{l_1,l_2,l_3,l_4}(e_p,e_\mathrm{per})$ are tabulated in Table \ref{table:ABD_n23} for $n=2$ and $n=3$ and the cases when $\mathcal{D}^{(n,i_1,i_2)}_{l_1,l_2,l_3,l_4}(e_p,e_\mathrm{per}) \neq 0$. The explicit expressions to the quadrupole-order ($n=2$) order are given by:
\begin{subequations}
\label{eq:delta_e_j_2}
\begin{align}
\label{eq:delta_e_j_2_e}
\nonumber &\Delta \ve{e}_p = \left( \frac{ M_\mathrm{per}^2}{M_p(M_p+M_\mathrm{per})} \right )^{1/2} \left ( \frac{a_p}{q_\mathrm{per}} \right )^{3/2} \frac{(1+e_\mathrm{per})^{-3/2}}{2e_\mathrm{per}^2} \Biggl [ \sqrt{e_\mathrm{per}^2-1} \Biggl \{ -6 \left ( \ve{e}_p \times \ve{\jmath}_p \right ) - 5 \left( \ve{e}_p \cdot \unit{\jmath}_\mathrm{per} \right ) \left ( \ve{\jmath}_p \times \unit{\jmath}_\mathrm{per} \right ) + 2  \left( \ve{\jmath}_p \cdot \unit{e}_\mathrm{per} \right ) \left ( \ve{e}_p \times \unit{e}_\mathrm{per} \right )  \\
\nonumber & \quad + \left( \ve{\jmath}_p \cdot \unit{\jmath}_\mathrm{per} \right ) \left ( \ve{e}_p \times \unit{\jmath}_\mathrm{per} \right ) + 10\left(e_\mathrm{per}^2-1\right) \left( \ve{e}_p \cdot \unit{e}_\mathrm{per} \right ) \left ( \ve{\jmath}_p \times \unit{e}_\mathrm{per} \right ) - 10 \,e_\mathrm{per}^2 \left( \ve{e}_p \cdot \unit{\jmath}_\mathrm{per} \right ) \left ( \ve{\jmath}_p \times \unit{\jmath}_\mathrm{per} \right ) - 2 \,e_\mathrm{per}^2 \left( \ve{\jmath}_p \cdot \unit{e}_\mathrm{per} \right ) \left ( \ve{e}_p \times \unit{e}_\mathrm{per} \right ) \\
&\quad + 2 \,e_\mathrm{per}^2 \left( \ve{\jmath}_p \cdot \unit{\jmath}_\mathrm{per} \right ) \left ( \ve{e}_p \times \unit{\jmath}_\mathrm{per} \right ) \Biggl \} + 3 \,e_\mathrm{per}^2 \, \mathrm{arcsec}(-e_\mathrm{per}) \biggl \{ -2 \left ( \ve{e}_p \times \ve{\jmath}_p \right ) - 5  \left( \ve{e}_p \cdot \unit{\jmath}_\mathrm{per} \right ) \left ( \ve{\jmath}_p \times \unit{\jmath}_\mathrm{per} \right ) +  \left( \ve{\jmath}_p \cdot \unit{\jmath}_\mathrm{per} \right ) \left ( \ve{e}_p \times \unit{\jmath}_\mathrm{per} \right ) \biggl \} \Biggl ]; 
\end{align}
\begin{align}
\nonumber &\Delta \ve{\jmath}_p = \left( \frac{ M_\mathrm{per}^2}{M_p(M_p+M_\mathrm{per})} \right )^{1/2} \left ( \frac{a_p}{q_\mathrm{per}} \right )^{3/2} \frac{(1+e_\mathrm{per})^{-3/2}}{2e_\mathrm{per}^2} \Biggl [ \sqrt{e_\mathrm{per}^2-1} \Biggl \{ - 5 \left( \ve{e}_p \cdot \unit{\jmath}_\mathrm{per} \right ) \left ( \ve{e}_p \times \unit{\jmath}_\mathrm{per} \right ) + 2  \left( \ve{\jmath}_p \cdot \unit{e}_\mathrm{per} \right ) \left ( \ve{\jmath}_p \times \unit{e}_\mathrm{per} \right )  + \\
\nonumber &\quad  \left( \ve{\jmath}_p \cdot \unit{\jmath}_\mathrm{per} \right ) \left ( \ve{\jmath}_p \times \unit{\jmath}_\mathrm{per} \right )  + 10\left(e_\mathrm{per}^2-1\right) \left( \ve{e}_p \cdot \unit{e}_\mathrm{per} \right ) \left ( \ve{e}_p \times \unit{e}_\mathrm{per} \right ) - 10 \, e_\mathrm{per}^2 \left( \ve{e}_p \cdot \unit{\jmath}_\mathrm{per} \right ) \left ( \ve{e}_p \times \unit{\jmath}_\mathrm{per} \right ) - 2 \,e_\mathrm{per}^2 \left( \ve{\jmath}_p \cdot \unit{e}_\mathrm{per} \right ) \left ( \ve{\jmath}_p \times \unit{e}_\mathrm{per} \right ) \\
&\quad + 2 \,e_\mathrm{per}^2 \left( \ve{\jmath}_p \cdot \unit{\jmath}_\mathrm{per} \right ) \left ( \ve{\jmath}_p \times \unit{\jmath}_\mathrm{per} \right ) \Biggl \} + 3 \,e_\mathrm{per}^2 \, \mathrm{arcsec}(-e_\mathrm{per}) \biggl \{   - 5  \left( \ve{e}_p \cdot \unit{\jmath}_\mathrm{per} \right ) \left ( \ve{e}_p \times \unit{\jmath}_\mathrm{per} \right ) +  \left( \ve{\jmath}_p \cdot \unit{\jmath}_\mathrm{per} \right ) \left ( \ve{\jmath}_p \times \unit{\jmath}_\mathrm{per} \right ) \biggl \} \Biggl ].
\end{align}
\end{subequations}
The scalar eccentricity change implied by equation~(\ref{eq:delta_e_j_2_e}) is
\begin{align}
\label{eq:delta_e_quad1}
\nonumber &\Delta e_p = \unit{e}_p \cdot \Delta \ve{e}_p = -\frac{5}{2} \frac{e_p}{e_\mathrm{per}^2} \left[ \frac{ M_\mathrm{per}^2}{M_p(M_p+M_\mathrm{per})}  \left ( \frac{a_p}{q_\mathrm{per}} \right )^3 \frac{1-e_p^2}{(1+e_\mathrm{per})^3} \right ]^{1/2} \Biggl [ \sqrt{e_\mathrm{per}^2-1} \Biggl \{  \left( \unit{e}_p \cdot \unit{\jmath}_\mathrm{per} \right ) \left [ \unit{e}_p \cdot \left ( \unit{\jmath}_p \times \unit{\jmath}_\mathrm{per} \right ) \right ] \\
&\quad - 2\left(e_\mathrm{per}^2-1\right) \left( \unit{e}_p \cdot \unit{e}_\mathrm{per} \right ) \left [ \unit{e}_p \cdot \left ( \unit{\jmath}_p \times \unit{e}_\mathrm{per} \right ) \right ] + 2 \,e_\mathrm{per}^2 \left( \unit{e}_p \cdot \unit{\jmath}_\mathrm{per} \right ) \left [ \unit{e}_p \cdot \left ( \unit{\jmath}_p \times \unit{\jmath}_\mathrm{per} \right ) \right ] \Biggl \} + 3 \,e_\mathrm{per}^2 \, \mathrm{arcsec}(-e_\mathrm{per}) \left( \unit{e}_p \cdot \unit{\jmath}_\mathrm{per} \right ) \left [ \unit{e}_p \cdot \left ( \unit{\jmath}_p \times \unit{\jmath}_\mathrm{per} \right ) \right ] \Biggl ].
\end{align}
We rewrite equation~(\ref{eq:delta_e_quad1}) to show that it is consistent with \citet{1996MNRAS.282.1064H}. Using a number of vector identities and changing the notation to that of \citet{1996MNRAS.282.1064H}, i.e., $\Delta e_p \rightarrow \delta e$, $e_p\rightarrow e$, $e_\mathrm{per} \rightarrow e'$, $a_p\rightarrow a$, $q_\mathrm{per} \rightarrow r_\mathrm{p}'$, $M_\mathrm{per} \rightarrow m_3$, $M_p\rightarrow M_{12}$, $M_p+M_\mathrm{per} \rightarrow M_{123}$, $\unit{e}_p \rightarrow \unit{a}$, $\unit{q}_p \rightarrow \unit{b}$, $\unit{e}_\mathrm{per} \rightarrow \unit{A}$ and $\unit{q}_\mathrm{per} \rightarrow \unit{B}$ (here, $\unit{q}_p \equiv \unit{\jmath}_p \times \unit{e}_p$ and $\unit{q}_\mathrm{per} \equiv \unit{\jmath}_\mathrm{per} \times \unit{e}_\mathrm{per}$), we find
\begin{align}
\label{eq:delta_e_quad2}
\nonumber \delta e &= -\frac{5}{2} \frac{e}{e'^2} \left[ \frac{ m_3^2}{M_{12} M_{123}}  \left ( \frac{a}{r'_\mathrm{p}} \right )^3 \frac{1-e^2}{(1+e')^3} \right ]^{1/2} \\
&\quad \times \Biggl \{ \left ( \unit{a} \cdot \unit{A}  \right ) \left ( \unit{b} \cdot \unit{A} \right ) \left [ 3 e'^2 \mathrm{arccos} \left ( -\frac{1}{e'} \right ) + \left (4e'^2-1 \right ) \sqrt{e'^2-1} \right ] + \left ( \unit{a} \cdot \unit{B}  \right ) \left ( \unit{b} \cdot \unit{B} \right ) \left [ 3 e'^2 \mathrm{arccos} \left ( -\frac{1}{e'} \right ) + \left (2e'^2+1 \right ) \sqrt{e'^2-1} \right ] \Biggl \}.
\end{align}
Equation~(\ref{eq:delta_e_quad2}) is identical to equation (A4) of \citet{1996MNRAS.282.1064H}.

\begin{table}
\def\arraystretch{2.5}
\begin{tabular}{ccccccccccl}
$n$ & $m$ & $i_1$ & $i_2$ & $l_1$ & $l_2$ & $l_3$ & $l_4$ & $\mathcal{A}_m^{(n)}$ & $\mathcal{B}_{i_1,i_2}^{(n,m)} (e_p)$ & $ \mathcal{D}^{(n,i_1,i_2)}_{l_1,l_2,l_3,l_4}(e_p,e_\mathrm{per})$ \\
\toprule
2& 0& 0& 0& 0& 0& 0& 0& $-\frac{1}{2}$& $1 + \frac{3}{2} e_p^2$ & 
  $2 \left [\sqrt{e_\mathrm{per}^2-1} + \mathrm{arcsec}(-e_\mathrm{per}) \right ]$ \\
2& 2& 0& 0& 0& 0& 0& 0& $\frac{3}{2}$& 
  $\frac{1}{2} \left(1 - e_p^2 \right)$ & $2 \left [ \sqrt{e_\mathrm{per}^2-1} + \mathrm{arcsec}(-e_\mathrm{per}) \right ]$ \\
2& 2& 0& 2& 0& 0& 0& 0& $\frac{3}{2}$ & $-\frac{1}{2}$& $\frac{1}{3} e_\mathrm{per}^{-2} \left (1 - e_p^2 \right ) \left [ \sqrt{e_\mathrm{per}^2-1} \left (1 + 2 e_\mathrm{per}^2\right) + 
      3 e_\mathrm{per}^2 \mathrm{arcsec}(-e_\mathrm{per}) \right ]$ \\
2& 2& 0& 2& 0& 0& 0& 2& $\frac{3}{2}$ & $-\frac{1}{2}$& $\frac{2}{3} e_\mathrm{per}^{-2} \left ( e_\mathrm{per}^2-1 \right )^{3/2}$ \\
2& 2& 0& 2& 0& 2& 0& 0& $\frac{3}{2}$ & $-\frac{1}{2}$& $-\frac{1}{3} e_\mathrm{per}^{-2} \left [ \sqrt{e_\mathrm{per}^2-1} \left(1 + 2 e_\mathrm{per}^2\right) + 3 e_\mathrm{per}^2 \mathrm{arcsec}(-e_\mathrm{per}) \right ]$ \\
2& 2& 2& 0& 0& 0& 0& 0& $\frac{3}{2}$& $\frac{5}{2}$& $\frac{1}{3} e_\mathrm{per}^{-2} e_p^2 \left [ \sqrt{e_\mathrm{per}^2-1} \left (1 + 2 e_\mathrm{per}^2\right ) + 3 e_\mathrm{per}^2 \mathrm{arcsec}(-e_\mathrm{per}) \right ]$ \\
2& 2& 2& 0& 0& 0& 2& 0& $\frac{3}{2}$& $\frac{5}{2}$& $-\frac{1}{3} e_\mathrm{per}^{-2} \left [ \sqrt{e_\mathrm{per}^2-1} \left (1 + 2 e_\mathrm{per}^2\right ) + 3 e_\mathrm{per}^2 \mathrm{arcsec}(-e_\mathrm{per}) \right ]$ \\
2& 2& 2& 0& 2& 0& 0& 0& $\frac{3}{2}$& $\frac{5}{2}$& $\frac{2}{3} e_\mathrm{per}^{-2} \left (e_\mathrm{per}^2-1 \right)^{3/2}$ \\
\midrule
3& 1& 1& 0& 1& 0& 0& 0& $-\frac{3}{2}$& $-\frac{5}{8} \left (4 + 3 e_p^2 \right )$& $\frac{1}{3} e_\mathrm{per}^{-1} \left [ 2 \sqrt{e_\mathrm{per}^2-1} \left(1 + 2 e_\mathrm{per}^2\right) + 6 e_\mathrm{per}^2 \mathrm{arcsec}(-e_\mathrm{per}) \right ]$ \\
3& 3& 1& 0& 1& 0& 0& 0& $\frac{5}{2}$& $-\frac{15}{8} \left (1 - e_p^2 \right)$ & $\frac{1}{3} e_\mathrm{per}^{-1} \left [ 2 \sqrt{e_\mathrm{per}^2-1} \left(1 + 2 e_\mathrm{per}^2\right) + 6 e_\mathrm{per}^2 \mathrm{arcsec}(-e_\mathrm{per}) \right ]$ \\
3& 3& 1& 2& 0& 1& 1& 1& $\frac{5}{2}$& $\frac{15}{8}$& $\frac{1}{15} e_\mathrm{per}^{-3} \left [ \sqrt{e_\mathrm{per}^2-1}  \left(2 - 9 e_\mathrm{per}^2 - 8 e_\mathrm{per}^4\right) - 15 e_\mathrm{per}^4 \mathrm{arcsec}(-e_\mathrm{per}) \right ]$ \\
3& 3& 1& 2& 1& 0& 0& 0& $\frac{5}{2}$& $\frac{15}{8}$& $-\frac{1}{30} \left(1-e_p^2\right) e_\mathrm{per}^{-3} \left [ \sqrt{e_\mathrm{per}^2-1}  \left(2 - 9 e_\mathrm{per}^2 - 8 e_\mathrm{per}^4\right) - 15 e_\mathrm{per}^4 \mathrm{arcsec}(-e_\mathrm{per}) \right ]$ \\
3& 3& 1& 2& 1& 0& 0& 2& $\frac{5}{2}$& $\frac{15}{8}$& $\frac{4}{15} e_\mathrm{per}^{-3} \left (e_\mathrm{per}^2-1\right)^{5/2}$ \\
3& 3& 1& 2& 1& 2& 0& 0& $\frac{5}{2}$ & $\frac{15}{8}$& $\frac{1}{30} e_\mathrm{per}^{-3} \left [ \sqrt{e_\mathrm{per}^2-1}  \left(2 - 9 e_\mathrm{per}^2 - 8 e_\mathrm{per}^4\right) - 15 e_\mathrm{per}^4 \mathrm{arcsec}(-e_\mathrm{per}) \right ]$ \\
3& 3& 3& 0& 1& 0& 0& 0& $\frac{5}{2}$& $-\frac{35}{8}$ & $-\frac{1}{10} e_p^2 e_\mathrm{per}^{-3} \left [ \sqrt{e_\mathrm{per}^2-1}  \left(2 - 9 e_\mathrm{per}^2 - 8 e_\mathrm{per}^4\right) - 15 e_\mathrm{per}^4 \mathrm{arcsec}(-e_\mathrm{per}) \right ]$ \\
3& 3& 3& 0& 1& 0& 2& 0& $\frac{5}{2}$& $-\frac{35}{8}$& $\frac{1}{10} e_\mathrm{per}^{-3} \left [ \sqrt{e_\mathrm{per}^2-1}  \left(2 - 9 e_\mathrm{per}^2 - 8 e_\mathrm{per}^4\right) - 15 e_\mathrm{per}^4 \mathrm{arcsec}(-e_\mathrm{per}) \right ]$ \\
3& 3& 3& 0& 3& 0& 0& 0& $\frac{5}{2}$& $-\frac{35}{8}$& $\frac{4}{15} e_\mathrm{per}^{-3} \left (e_\mathrm{per}^2-1\right)^{5/2}$ \\
\bottomrule
\end{tabular}
\caption{ The coefficient $\mathcal{A}_m^{(n)}$ and the functions $\mathcal{B}_{i_1,i_2}^{(n,m)} (e_p)$ and $ \mathcal{D}^{(n,i_1,i_2)}_{l_1,l_2,l_3,l_4}(e_p,e_\mathrm{per})$ for $n=2$ and $n=3$ and the cases when $\mathcal{D}^{(n,i_1,i_2)}_{l_1,l_2,l_3,l_4}(e_p,e_\mathrm{per}) \neq 0$.}
\label{table:ABD_n23}
\end{table}

\newpage

\section{Code Fragments}

\begin{lstlisting}[caption={Illustration of code usage within \textsc{Python} and \textsc{AMUSE} to compute the effect of a secular encounter with a binary (see also \S\,\ref{sect:sec:code}). Here, the system is initialized. See Code Fragments \ref{code:sec:app1} and \ref{code:sec:app2} for the usage of approaches (1) and (2), respectively. },label={code:sec:init},language=Python]
import numpy
numpy.random.seed(0)

from amuse.community.secularmultiple.interface import SecularMultiple
from amuse.units import quantities,units,constants
from amuse.datamodel import Particles

### Set up the hierarchical system ###
N_bodies = 2
N_binaries = N_bodies-1

masses = [1.0|units.MSun, 0.8|units.MSun]
semimajor_axes = [1.0|units.AU]
eccentricities = [0.1]
inclinations = [numpy.random.rand()*numpy.pi for i in range(N_binaries)]
arguments_of_pericentre = [numpy.random.rand()*2.0*numpy.pi for i in range(N_binaries)]
longitudes_of_ascending_node = [numpy.random.rand()*2.0*numpy.pi for i in range(N_binaries)]

particles = Particles(N_bodies+N_binaries)
for index in range(N_bodies):
    particle = particles[index]
    particle.mass = masses[index]
    particle.is_binary = False
    particle.radius = 1.0 | units.RSun
    particle.child1 = None
    particle.child2 = None

for index in range(N_binaries):
    particle = particles[index+N_bodies]
    particle.is_binary = True
    particle.semimajor_axis = semimajor_axes[index]
    particle.eccentricity = eccentricities[index]
    particle.inclination = inclinations[index]
    particle.argument_of_pericenter = arguments_of_pericentre[index]
    particle.longitude_of_ascending_node = longitudes_of_ascending_node[index]

    if index==0:
        particle.child1 = particles[0]
        particle.child2 = particles[1]
    elif index==1:
        particle.child1 = particles[2]
        particle.child2 = particles[3]
binaries = particles[particles.is_binary]

### Specify the perturber ###
external_particles = Particles(1)
external_particle = external_particles[0]

q_per, M_per = 100.0 | units.AU, 1.0 | units.MSun
t_char = numpy.pi*numpy.sqrt( q_per**3/(2.0*constants.G*(M_per+masses[0]+masses[1])) )
external_particle.mass = M_per
external_particle.periapse_distance = q_per     
external_particle.eccentricity = 5.0
external_particle.t_ref = t_char

external_particle.e_hat_vec_x = 1.0
external_particle.e_hat_vec_y = 0.0
external_particle.e_hat_vec_z = 0.0

external_particle.h_hat_vec_x = 0.0
external_particle.h_hat_vec_y = 0.0
external_particle.h_hat_vec_z = 1.0
\end{lstlisting}

\begin{lstlisting}[caption={The example of Code Fragment \ref{code:sec:init} continued, here adoping approach (1); see Code Fragment \ref{code:sec:app2} for approach (2). },label={code:sec:app1},language=Python]
### Approach (1): numerically integrate over the perturber orbit ###
external_particle.path = 1 ### 1 for hyperbolic orbit (straight line: 0)
external_particle.mode = 0 ### 0 for approach 1

channel_from_particles_to_code.copy()
channel_from_external_particles_to_code.copy()

time, end_time, output_time_step = 0.0|units.yr, 2.0*t_char, t_char/100.0
while time <= end_time:
    time += output_time_step
    code.evolve_model(time)

    channel_from_code_to_particles.copy()
    print '='*50
    print 't/yr',time.value_in(units.yr)
    print 'e',binaries.eccentricity
    print 'i/deg',binaries.inclination*(180.0/numpy.pi)  
    print 'AP/deg', binaries.argument_of_pericenter*(180.0/numpy.pi)  
    print 'LAN/deg', binaries.longitude_of_ascending_node*(180.0/numpy.pi)
\end{lstlisting}

\begin{lstlisting}[caption={The example of Code Fragment \ref{code:sec:init} continued, here adoping approach (2); see Code Fragment \ref{code:sec:app1} for approach (1). },label={code:sec:app2},language=Python]
### Approach (2): analytically integrate over the perturber orbit ###
external_particle.path = 1 ### 1 for hyperbolic orbit (straight line: 0, not supported for approach 2)
external_particle.mode = 1 ### 1 for approach 2

channel_from_particles_to_code.copy()
channel_from_external_particles_to_code.copy()

code.apply_external_perturbation_assuming_integrated_orbits()
channel_from_code_to_particles.copy()

print 'e',binaries.eccentricity
print 'i/deg',binaries.inclination*(180.0/numpy.pi)  
print 'AP/deg', binaries.argument_of_pericenter*(180.0/numpy.pi)  
print 'LAN/deg', binaries.longitude_of_ascending_node*(180.0/numpy.pi)
\end{lstlisting}

\begin{lstlisting}[caption={Illustration of code usage to compute the effect of an asymmetric SN on a hierarchical sextuple system (see also \S\,\ref{sect:inst:impl:use}).},label={code:inst},language=Python]
import numpy
numpy.random.seed(0)

from amuse.community.secularmultiple.interface import SecularMultiple
from amuse.units import quantities,units,constants
from amuse.datamodel import Particles

### Set up the hierarchical system ###
N_bodies = 6
N_binaries = N_bodies-1

masses = [15.0|units.MSun, 8.1|units.MSun, 12.8|units.MSun, 11.2 | units.MSun,12.2 | units.MSun,10.2 | units.MSun]
semimajor_axes = [10.0|units.AU, 5.0|units.AU,100.0|units.AU, 20|units.AU, 1000.0|units.AU]
eccentricities = [0.1,0.2,0.3,0.1,0.4]
inclinations = [numpy.random.rand()*numpy.pi for i in range(N_binaries)]
arguments_of_pericentre = [numpy.random.rand()*2.0*numpy.pi for i in range(N_binaries)]
longitudes_of_ascending_node = [numpy.random.rand()*2.0*numpy.pi for i in range(N_binaries)]
true_anomalies = [numpy.random.rand()*2.0*numpy.pi for i in range(N_binaries)]
sample_orbital_phases_randomly = False

particles = Particles(N_bodies+N_binaries)
for index in range(N_bodies):
    particle = particles[index]
    particle.mass = masses[index]
    particle.is_binary = False
    particle.radius = 1.0 | units.RSun
    particle.child1 = None
    particle.child2 = None

for index in range(N_binaries):
    particle = particles[index+N_bodies]
    particle.is_binary = True
    particle.semimajor_axis = semimajor_axes[index]
    particle.eccentricity = eccentricities[index]
    particle.true_anomaly = true_anomalies[index]
    particle.inclination = inclinations[index]
    particle.argument_of_pericenter = arguments_of_pericentre[index]
    particle.longitude_of_ascending_node = longitudes_of_ascending_node[index]
    particle.sample_orbital_phases_randomly = sample_orbital_phases_randomly

    if index==0:
        particle.child1 = particles[0]
        particle.child2 = particles[1]
    elif index==1:
        particle.child1 = particles[2]
        particle.child2 = particles[3]
    elif index==2:
        particle.child1 = particles[6]
        particle.child2 = particles[7]
    elif index==3:
        particle.child1 = particles[4]
        particle.child2 = particles[5]
    elif index==4:
        particle.child1 = particles[8]
        particle.child2 = particles[9]
        
### Specify the instantaneous changes ###
particles[0].instantaneous_perturbation_delta_mass = -4.0 | units.MSun
particles[0].instantaneous_perturbation_delta_velocity_x = 10.0 | units.km/units.s
particles[0].instantaneous_perturbation_delta_velocity_y = 2.0 | units.km/units.s
particles[0].instantaneous_perturbation_delta_velocity_z = 3.0 | units.km/units.s

### Initialize the code, add the particles, and set up channels ###
code = SecularMultiple()
code.particles.add_particles(particles)

channel_from_particles_to_code = particles.new_channel_to(code.particles)
channel_from_code_to_particles = code.particles.new_channel_to(particles)

binaries = particles[particles.is_binary]
binaries.sample_orbital_phases_randomly = False ### set to "True" to sample all orbital phases randomly
code.parameters.orbital_phases_random_seed = 0

channel_from_particles_to_code.copy()

### Print the pre-SN state ###
print '='*50
print 'pre SN'
print 'a/AU',(binaries.semimajor_axis)
print 'e',(binaries.eccentricity)
print 'TA/deg',binaries.true_anomaly*(180.0/numpy.pi)  
print 'i/deg',binaries.inclination*(180.0/numpy.pi)  
print 'AP/deg', binaries.argument_of_pericenter*(180.0/numpy.pi)  
print 'LAN/deg', binaries.longitude_of_ascending_node*(180.0/numpy.pi)

### Compute the effect of the SN on the system and copy to user-level particles ###
code.apply_user_specified_instantaneous_perturbation()
channel_from_code_to_particles.copy()

### Print the post-SN state ###
print '='*50
print 'post SN'
print 'a/deg', (binaries.semimajor_axis)
print 'e', (binaries.eccentricity)
print 'TA/deg',binaries.true_anomaly*(180.0/numpy.pi)  
print 'i/deg',binaries.inclination*(180.0/numpy.pi)  
print 'AP/deg', binaries.argument_of_pericenter*(180.0/numpy.pi)  
print 'LAN/deg', binaries.longitude_of_ascending_node*(180.0/numpy.pi)
\end{lstlisting}

\begin{lstlisting}[caption={Illustration of code usage to compute the effect of an impulsive encounter with a hierarchical quadruple system (see also \S\,\ref{sect:inst:imp:code}).},label={code:inst_imp},language=Python]
import numpy
numpy.random.seed(0)

from amuse.community.secularmultiple.interface import SecularMultiple
from amuse.units import quantities,units,constants
from amuse.datamodel import Particles

### Set up the hierarchical system ###
N_bodies = 4
N_binaries = N_bodies-1

masses = [10.0|units.MSun, 8.1|units.MSun, 12.8|units.MSun, 11.2 | units.MSun]
semimajor_axes = [10.0|units.AU, 50.0|units.AU,5000.0|units.AU]
eccentricities = [0.1,0.2,0.3]
inclinations = [numpy.random.rand()*numpy.pi for i in range(N_binaries)]
arguments_of_pericentre = [numpy.random.rand()*2.0*numpy.pi for i in range(N_binaries)]
longitudes_of_ascending_node = [numpy.random.rand()*2.0*numpy.pi for i in range(N_binaries)]
sample_orbital_phases_randomly = True

particles = Particles(N_bodies+N_binaries)
for index in range(N_bodies):
    particle = particles[index]
    particle.mass = masses[index]
    particle.is_binary = False
    particle.radius = 1.0 | units.RSun
    particle.child1 = None
    particle.child2 = None

for index in range(N_binaries):
    particle = particles[index+N_bodies]
    particle.is_binary = True
    particle.semimajor_axis = semimajor_axes[index]
    particle.eccentricity = eccentricities[index]
    particle.inclination = inclinations[index]
    particle.argument_of_pericenter = arguments_of_pericentre[index]
    particle.longitude_of_ascending_node = longitudes_of_ascending_node[index]
    particle.sample_orbital_phases_randomly = sample_orbital_phases_randomly
    
    if index==0:
        particle.child1 = particles[0]
        particle.child2 = particles[1]
    elif index==1:
        particle.child1 = particles[2]
        particle.child2 = particles[3]
    elif index==2:
        particle.child1 = particles[4]
        particle.child2 = particles[5]        

binaries = particles[particles.is_binary]
bodies = particles-binaries

code = SecularMultiple(redirection='none')
code.particles.add_particles(particles)

channel_from_particles_to_code = particles.new_channel_to(code.particles)
channel_from_code_to_particles = code.particles.new_channel_to(particles)
channel_from_particles_to_code.copy()
channel_from_code_to_particles.copy()

### Specify the (impulsive) perturber ###
M_per = 1.0 | units.MSun
b = 1.0e2 | units.AU
b_vec_unit = numpy.array([0.0,0.0,1.0])
b_vec = b*b_vec_unit
V_per = 50.0 | units.km/units.s
V_per_unit = numpy.array([0.0,1.0,0.0])

particles.add_vector_attribute("position", ["position_x","position_y","position_z"])

for index,particle in enumerate(bodies):
    R_vec = particle.position
    b_i_vec = b_vec - R_vec - V_per_unit*(b_vec - R_vec).dot(V_per_unit)
    Delta_V_vec = 2.0*(constants.G*M_per/V_per)*b_i_vec/(b_i_vec.lengths_squared())
    
    particles[index].instantaneous_perturbation_delta_velocity_x = Delta_V_vec[0]
    particles[index].instantaneous_perturbation_delta_velocity_y = Delta_V_vec[1]
    particles[index].instantaneous_perturbation_delta_velocity_z = Delta_V_vec[2] 

### Print the pre-encounter state ###
print '='*50
print 'pre SN'
print 'a/AU',(binaries.semimajor_axis)
print 'e',(binaries.eccentricity)
print 'TA/deg',binaries.true_anomaly*(180.0/numpy.pi)  
print 'i/deg',binaries.inclination*(180.0/numpy.pi)  
print 'AP/deg', binaries.argument_of_pericenter*(180.0/numpy.pi)  
print 'LAN/deg', binaries.longitude_of_ascending_node*(180.0/numpy.pi)

### Copy the Delta V_i's to particles in the code, and compute the orbital changes ###
channel_from_particles_to_code.copy()
code.apply_user_specified_instantaneous_perturbation()
channel_from_code_to_particles.copy()

### Print the post-encounter state ###
print '='*50
print 'post SN'
print 'a/deg', (binaries.semimajor_axis)
print 'e', (binaries.eccentricity)
print 'TA/deg',binaries.true_anomaly*(180.0/numpy.pi)  
print 'i/deg',binaries.inclination*(180.0/numpy.pi)  
print 'AP/deg', binaries.argument_of_pericenter*(180.0/numpy.pi)  
print 'LAN/deg', binaries.longitude_of_ascending_node*(180.0/numpy.pi)
\end{lstlisting}

\label{lastpage}

\end{document}